\documentclass[epj]{svjour}
\usepackage{latexsym}
\usepackage{graphics}
\usepackage{amsmath}
\usepackage[utf8]{inputenc}
\usepackage{morefloats}
\usepackage{hyperref}
\begin{document}
\title{Wikipedia Ranking of World Universities}
\author{
José Lages\inst{1}\thanks{\emph{email address:} jose.lages@utinam.cnrs.fr}
\and
Antoine Patt\inst{1}
\and
Dima L. Shepelyansky\inst{2}\thanks{\emph{email address:} dima@irsamc.ups-tlse.fr}
}                    
\institute{
Institut UTINAM, Observatoire des Sciences de l'Univers THETA, CNRS, 
Université de Franche-Comté, 25030 Besançon, France
\and
Laboratoire de Physique Théorique du CNRS, IRSAMC, Université de 
Toulouse, UPS, 31062 Toulouse, France
}
\date{Received: 29 November 2015 / Revised version: date}
\abstract{
We use the directed networks between articles of 24 Wikipedia language editions
for producing the Wikipedia Ranking of World Universities (WRWU) using 
PageRank, 2DRank and CheiRank algorithms. This approach 
allows to incorporate various cultural views on world universities
using the mathematical statistical analysis 
independent of cultural preferences.  
The Wikipedia ranking of top 100 universities provides 
about 60 percent overlap with the Shanghai university ranking
demonstrating the reliable features of this approach.
At the same time WRWU incorporates all knowledge accumulated 
at 24 Wikipedia editions giving stronger highlights
for historically important universities leading to a
different estimation of efficiency of world countries in
university education. The historical development of university
ranking is analyzed during ten centuries of their history.
 % end of PACS codes
} %end of abstract
\maketitle
\section{Introduction}
\label{intro}
According to the UNESCO reports the higher education is definitely at the heart of 
modern society development and related academic revolution
(see e.g. \cite{unesco2009}). Thus the analysis of the efficiency
of university education in different countries
becomes of political importance for the country future development.
One of the important tools of this analysis is the 
university ranking reviewed in high details
at \cite{hazelkorn}. Indeed, it is now
well established that the Academic Ranking of World
Universities (ARWU), 
compiled by Shanghai Jiao Tong University since 2003 (Shanghai
ranking) \cite{shanghai}, produced a significant
impact on evaluation of national universities
both on educational and political levels \cite{unesco2009,hazelkorn}.
Thus, for example, ARWU affected the French
strategies LABEX, IDEX in high education
\cite{freducation}.
Also the Russian Academic Excellence Project 
with significant financial investments \cite{ru5top100} 
in many respects has been initiated by ARWU. 
Other examples are reviewed in \cite{hazelkorn}.
At present there are several additional university rankings 
which are based on various evaluation methods
of university efficiency in research and education (see e.g. 
\cite{times,umultirank,ireg}).

The scientific analysis of strong and weak features of 
various university ranking methods is performed by
various research groups  as reported for example in
\cite{docampo2011,bornmann2013,jons2013,docampo2014}.
A comparative analysis of various approaches
is given in \cite{hazelkorn,eua2013}. It is in general accepted that
the world university rankings
play an important role for development 
of higher education in the world countries,
even if there are various opinions about each approach.

The above scientific studies definitely show the importance
of university ranking. These ranking approaches are based
on human selection rules which can not be complete or can favor 
certain cultural choices and preferences.
Thus it is useful to have an
independent mathematical statistical method which would 
rank universities independently of any human
rules. Such a method has been proposed in  \cite{wikizzs}
being based on the mathematical analysis of the
human knowledge accumulated at English Wikipedia by year 2009
(www.wikipedia.org). This approach is based 
on a  directed 
network of citations between all available 
articles of Wikipedia, construction of the 
corresponding Markov chain transitions
\cite{markov}  
and the Google matrix $G$, introduced by
Brin and Page in 1998 \cite{brin}
for hypertext analysis of the World Wide Web (WWW).
The construction rules of $G$ matrix
and description of its spectral properties
for various directed networks are given 
in \cite{meyer,rmpefs}. 
The general scale-free properties of
complex networks are described in \cite{dorogovtsev}.
The studies performed in \cite{wikizzs,wikievol}
demonstrated that this approach recovers
about 70\% and 80\% of top 100 and top 10 universities of ARWU
and that this overlap remains stable during the time
evolution of English Wikipedia during the years
$2004 - 2011$.

A similar approach based on the Wikipedia network
was used for ranking of historical figures 
of English Wikipedia \cite{wikizzs,wikievol}.
The extension of this approach to 9 \cite{eomwiki9}
and 24 language editions of Wikipedia \cite{eomwiki24}
allowed to take into account various cultural view points
and improve the overlap of top 100 historical 
figures from Wikipedia with the Hart top 100 people, who
according to him, most influenced human history \cite{hart}.
The approaches of different groups to
the Wikipedia ranking of historical figures are
discussed in \cite{wikizzs,eomwiki24,dlswiki100}.
The results for the top 100 historical figures of Wikipedia
approve the validity of this mathematical ranking approach
based on human knowledge accumulated in 
various language editions of Wikipedia.

In this work we extend this approach 
creating the Wikipedia Ranking of World Universities (WRWU).
We use the network data set of 24 Wikipedia language editions
collected at  \cite{eomwiki24}. These 24 languages cover
59\% of world population and 68\% of the total number
of Wikipedia articles in all 287 languages.
On the basis of the developed analysis we 
determine the most influential 
universities in the world and 
consider their time and geographical evolution on
a scale of 10 centuries of human history.
This study also allows to consider the
various cultural preferences in the
importance of concrete universities 
by different countries. Our WRWU 
results have about 60\% and 90\% 
overlap with the top 100 and top 10 list of ARWU.

The paper is constructed as follows:
In Section 2, we describe Wikipedia data sets used in this work and 
we introduce the WRWU approach
which is based on the Google matrix and PageRank, CheiRank, 
2DRank algorithms. In Section 3, results of WRWU are compared to ARWU with 
a geographical and temporal analysis. In Section 4, we study entanglement 
of cultures and their interactions through WRWU results. 
Finally, the discussion of the results is
given in Section 5.

\section{Description of data sets and methods}

We consider 24 Wikipedia language editions already used 
to rank historical figures of Wikipedia 
\cite{eomwiki24}:
Arabic (AR),
Danish (DA),
German (DE),
Greek (EL),
English (EN),
Spanish (ES),
Persian (FA),
French (FR),
Hebrew (HE),
Hindi (HI),
Hungarian (HU),
Italian (IT),
Japanese (JA),
Korean (KO),
Malaysian (MS),
Dutch (NL),
Polish (PL),
Portuguese (PT),
Russian (RU),
Swedish (SV),
Thai (TH),
Turkish (TR),
Vietnamese (VI),
Chinese (ZH).
Titles of Wikipedia articles and hyperlinks between articles 
were collected in middle February 2013 (see \cite{eomwiki24} for data preparation details).

\subsection{Network definition}
Following \cite{eomwiki24}, we consider each of the Wikipedia language editions as 
an isolated directed network whose  nodes are articles and 
the directed links  are formed by citations from one article to another article. 
In this study we do not consider 
hyperlinks between different language editions. 
We associate to a given network an adjacency matrix $A$ with elements $A_{ij}$ 
being $1$ if node (article) $j$ points towards node (article) $i$ and $0$ otherwise. 
A network associated to a given Wikipedia language edition containing 
$N$ articles connected with $N_\ell$ hyperlinks is then characterized 
by its $N \times N$ adjacency matrix $A$ containing $N_\ell$ non zero $A_{ij}$ elements. 
The parameters  of the networks constructed from  24 Wikipedia language editions
are given in Table~\ref{table1} (see also \cite{eomwiki24}). 
The country codes (CC)
and language codes (LC) are given in Table~\ref{table2}. The
CC codes follow ISO 3166-1 alpha-2 standard \cite{isowiki} and the
LC codes are language edition codes of Wikipedia, the
code WR represents all languages other than 
the considered 24 languages.

\begin{table}
\caption{
Wikipedia directed networks from  24 considered language editions;
here $N$ is the number of articles. Wikipedia data were collected in
middle February 2013 \cite{eomwiki24}.}
\resizebox{\columnwidth}{!}{
\begin{tabular}{llrllr}
\hline
Edition&Language&$N$&Edition&Language&$N$\\
\hline
EN &English& 4212493&VI &Vietnamese &594089\\
DE &German& 1532978&FA &Persian& 295696\\
FR &French& 1352825&HU &Hungarian& 235212\\
NL &Dutch& 1144615&KO &Korean& 231959\\
IT &Italian& 1017953&TR &Turkish& 206311\\
ES &Spanish& 974025&AR &Arabic& 203328\\
RU &Russian& 966284&MS &Malaysian& 180886\\
PL &Polish& 949153&DA &Danish& 175228\\
JA &Japanese& 852087&HE &Hebrew& 144959\\
SV &Swedish& 780872&HI &Hindi& 96869\\
PT &Portuguese& 758227&EL &Greek& 82563\\
ZH &Chinese& 663485&TH &Thai& 78953\\
%AR &Arabic &203328&JA &Japanese &852087\\
%DA &Danish &175228&KO &Korean &231959\\
%DE &German &1532978&MS &Malaysian &180886\\
%EL &Greek &82563&NL &Dutch &1144615\\
%EN &English &4212493&PL &Polish &949153\\
%ES &Spanish &974025&PT &Portuguese &758227\\
%FA &Persian &295696&RU &Russian &966284\\
%FR &French &1352825&SV &Swedish &780872\\
%HE &Hebrew &144959&TH &Thai &78953\\
%HI &Hindi &96869&TR &Turkish &206311\\
%HU &Hungarian &235212&VI &Vietnamese &594089\\
%IT &Italian &1017953&ZH &Chinese &663485\\ 
\hline
\end{tabular}
}
\label{table1}
\end{table}

\begin{table*}
\caption{
List of countries with corresponding country codes (CC) and language 
codes (LC).
Only countries appearing in the top 100 universities of  24 Wikipedia 
editions
using PageRank, CheiRank, and 2DRank algorithms are listed here. LC is 
determined by
the most spoken language in the given country.
Country codes (CC) follow ISO 3166-1 alpha-2 standard \cite{isowiki}.
Language codes are based on language edition codes of Wikipedia;
WR represents all languages other than the considered 24 languages.
The data are represented by three columns with CC Country LC.}
\centering
%\resizebox{\columnwidth}{!}{
\begin{tabular}{lll|lll|lll}
\hline
CC&Country&LC&CC&Country&LC&CC&Country&LC\\
\hline
AE & United Arab Emirates & AR & GU & Guam & EN & OM & Oman & AR\\
AF & Afghanistan & FA & GY & Guyana & EN & PA & Panama & ES\\
AL & Albania & WR & HK & Hong Kong & ZH & PE & Peru & ES\\
AM & Armenia & WR & HN & Honduras & ES & PG & Papua New Guinea & EN\\
AO & Angola & PT & HR & Croatia & WR & PH & Philippines & EN\\
AR & Argentina & ES & HT & Haiti & FR & PK & Pakistan & HI\\
AT & Austria & DE & HU & Hungary & HU & PL & Poland & PL\\
AU & Australia & EN & ID & Indonesia & WR & PR & Puerto Rico & ES\\
AZ & Azerbaijan & TR & IE & Ireland & EN & PS & State of Palestine & AR\\
BD & Bangladesh & WR & IL & Israel & HE & PT & Portugal & PT\\
BE & Belgium & NL & IN & India & HI & PY & Paraguay & ES\\
BF & Burkina Faso & FR & IQ & Iraq & AR & QA & Qatar & AR\\
BG & Bulgaria & WR & IR & Iran & FA & RO & Romania & WR\\
BH & Bahrain & AR & IS & Iceland & WR & RS & Serbia & WR\\
BJ & Benin & FR & IT & Italy & IT & RU & Russia & RU\\
BN & Brunei & MS & JM & Jamaica & EN & RW & Rwanda & EN\\
BR & Brazil & PT & JO & Jordan & AR & SA & Saudi Arabia & AR\\
BS & Bahamas & EN & JP & Japan & JA & SD & Sudan & AR\\
BT & Bhutan & WR & KE & Kenya & EN & SE & Sweden & SV\\
BY & Belarus & RU & KG & Kyrgyzstan & WR & SG & Singapore & ZH\\
CA & Canada & EN & KH & Cambodia & WR & SI & Slovenia & WR\\
CF & Central African Republic & FR & KM & Comoros & FR & SK & Slovakia & 
WR\\
CH & Switzerland & DE & KP & North Korea & KO & SO & Somalia & WR\\
CI & Ivory Coast & FR & KR & South Korea & KO & SR & Suriname & NL\\
CL & Chile & ES & KW & Kuwait & AR & SV & El Salvador & ES\\
CN & China & ZH & KZ & Kazakhstan & WR & SY & Syria & AR\\
CO & Colombia & ES & LA & Laos & WR & SZ & Swaziland & EN\\
CR & Costa Rica & ES & LB & Lebanon & AR & TH & Thailand & TH\\
CU & Cuba & ES & LK & Sri Lanka & WR & TJ & Tajikistan & WR\\
CY & Cyprus & EL & LR & Liberia & EN & TL & Timor-Leste & PT\\
CZ & Czech Republic & WR & LT & Lithuania & WR & TN & Tunisia & AR\\
DE & Germany & DE & LV & Latvia & WR & TR & Turkey & TR\\
DK & Denmark & DA & LY & Libya & AR & TW & Taiwan & ZH\\
DO & Dominican Republic & ES & MA & Morocco & AR & TZ & Tanzania & WR\\
DZ & Algeria & AR & MC & Monaco & FR & UA & Ukraine & WR\\
EC & Ecuador & ES & MD & Moldova & WR & UG & Uganda & EN\\
EE & Estonia & WR & MK & Macedonia & WR & UK & United Kingdom & EN\\
EG & Egypt & AR & MM & Myanmar & WR & US & United States & EN\\
ES & Spain & ES & MN & Mongolia & WR & UY & Uruguay & ES\\
ET & Ethiopia & EN & MT & Malta & EN & UZ & Uzbekistan & WR\\
FI & Finland & WR & MW & Malawi & EN & VA & Holy See & IT\\
FJ & Fiji & EN & MX & Mexico & ES & VE & Venezuela & ES\\
FO & Faroe Islands & DA & MY & Malaysia & MS & VN & Vietnam & VI\\
FR & France & FR & NG & Nigeria & EN & YE & Yemen & AR\\
GE & Georgia & WR & NL & Netherlands & NL & ZA & South Africa & WR\\
GH & Ghana & EN & NO & Norway & WR & ZW & Zimbabwe & EN\\
GL & Greenland & DA & NP & Nepal & WR & &  & \\
GR & Greece & EL & NZ & New Zealand & EN & &  & \\
\hline
\end{tabular}
%}
\label{table2}
\end{table*}

\subsection{Google matrix}

We suppose that a random surfer hops from a node $j$ to 
any connected node $i$ ($A_{ij}=1$) with probability $1/k_{out}(j)$ 
where $k_{out}(j)=\sum_{i=1}^{N}A_{ij}\neq0$ is the node $j$ out-degree, 
\textit{i.e.} the number of links from node $j$ to other nodes. 
If node $j$ is a dangling node without outgoing links ($k_{out}(j)=0$), 
then we assume that a random surfer hops to any of the  network 
nodes $N$ with the probability $1/N$.
Then the  matrix of Markov transitions
$S$ is defined by its elements 
$S_{ij}=A_{ij}/k_{out}(j)$ if $k_{out}(j)\neq0$ and $S_{ij}=1/N$ otherwise.
The Google matrix $G$ is defined by the standard relation \cite{brin,meyer}:
\begin{equation}
  G_{ij} = \alpha S_{ij} + (1-\alpha) / N \;\; ,
\label{eq:gmatrix} 
\end{equation}
where $\alpha$ is the  damping factor. 
We use throughout the paper the conventional value $\alpha=0.85$. 
The values $\alpha$  in the range $0.5 \leq \alpha <0.95$ 
do not affect the ranking  \cite{meyer,rmpefs,eomwiki24}.

\subsection{PageRank, CheiRank and 2DRank algorithms}

The PageRank algorithm \cite{brin,meyer} allows to rank all nodes of the network.
Let us assume that a random surfer journey starts from node $k$. 
We define $P_i(t)$ the probability that a random surfer reaches node $i$ after $t$ 
iterations with $P_k(0)=1$ and $P_i(0)=0$ for $i\neq k$. 
The probability vector $\mathbf{P}(t)$ whose the components 
are the probabilities $P_i(t)$ is given by
\begin{equation}\label{eq:PR}
\mathbf{P}(t)=\underset{\mbox{$t$ times}}
{\underbrace{GG\dots G}}\,\mathbf{P}(0)=G^t\mathbf{P}(0).
\end{equation}
Providing $\alpha<1$, for any given $\mathbf{P}(0)$, 
the probability vector $\mathbf{P}(t)$ converges towards 
an unique stationary vector $\mathbf{P}$ as the number of iterations increases. 
This is the right eigenvector of $G$ matrix 
with the eigenvalue $\lambda=1$ ($G \mathbf{P} = \lambda \mathbf{P}$).
In our numerical simulations we compute iteratively 
$\mathbf{P}$ up to a precision of $10^{-17}$, 
\textit{i.e.} we compute $P(t)$ up to iteration $t'$ such
as $\sum_{i=1}^N\arrowvert P_i(t')-P_i(t'-1)\arrowvert \leq 10^{-17}$. 
The $i$th component of $\mathbf P$, $P_i$, gives the average proportion of 
time spent by a random surfer on node $i$. 
Ordering the probabilities $P_i$ from biggest to smallest gives 
the PageRank \cite{brin} index $K$ with $K=1$ ($K=N$) 
associated to node with maximum (minimum) probability. 

It is also useful to consider the network with inverted direction of links. 
Then the matrix of Markov transitions is  defined as $S^*_{ij}=A_{ji}/k_{in}(j)$ 
if $k_{in}(j)\neq0$ and $S_{ij}=1/N$ otherwise. 
Here $k_{in}(j)=\sum_{j=1}^{N}A_{ij}$ is the node $j$ in-degree 
\textit{i.e.} the number of links to node $j$ from other nodes. 
The associated dual Google matrix $G^*$ is consequently $G^*_{ij}=\alpha S^*_{ij}+(1-\alpha)/N$. 
Similarly to PageRank, it is possible to define a probability vector
\begin{equation}
\mathbf{P}^*(t)=\underset{\mbox{$t$ times}}
{\underbrace{G^*G^*\dots G^*}}\,\mathbf{P}^*(0)={G^*}^t\mathbf{P}^*(0).
\end{equation}
which, providing $\alpha<1$, for any given $\mathbf{P}^*(0)$, 
converges towards an unique stationary probability vector $\mathbf{P}^*$. 
Probability $P^*_i$ gives the average time spent on node $i$ 
by a random surfer evolving on the inverted directed network. 
The probability vector $\mathbf{P}^*$ is the right eigenvector
of the matrix $G^*$ with the eigenvalue $\lambda=1$
($G^* \mathbf{P}^* = \lambda \mathbf{P}^*$).
The statistical properties of this CheiRank vector
have been analyzed in \cite{linux} (see also \cite{wikizzs,rmpefs}).
By ordering the probabilities $P^*_i$ from largest to smallest values
gives the CheiRank index with   $K^*=1$ ($K^*=N$) 
associated to node with maximum (minimum) probability.

The probability of PageRank vector is proportional to 
the number of ingoing links
while the probability of the CheiRank vector is proportional 
to the number of outgoing links
(see e.g. \cite{meyer,rmpefs}).
It is also possible to define a third ranking, 2DRank, 
which combines PageRank and CheiRank \cite{wikizzs}. 
Assuming a node with PageRank $K$ and CheiRank $K^*$, 
the 2DRank index for this node is $K_2=\max\{K,K^*\}$. 
The 2DRank algorithm and 2DRank index $K_2$ 
are described in detail  at \cite{wikizzs}. 
Thus the PageRank index $K$ have at the top
well known articles of Wikipedia (e.g. world countries)
while the CheiRank index $K^*$ has at the top
very communicative article 
(e.g. listings of geographical names, prime ministers etc.).
The top articles of 2DRank index $K_2$ are those 
which are both well known and communicative (see \cite{wikizzs,rmpefs,linux}).
We note that PageRank and CheiRank appear very naturally
in the trade networks corresponding to import and export flows \cite{rmpefs}. 
For the Wikipedia networks the global properties of PageRank, CheiRank
and 2DRank have been discussed in detail in 
\cite{wikizzs,wikievol,eomwiki9,eomwiki24}.

\subsection{Rankings of world universities}

For each individual Wikipedia language edition 
we rank all $N$ articles using PageRank, 
CheiRank, and 2DRank algorithms. We consequently obtain three different
global rank indexes
$K, K^*, K_2$
from which we extract articles devoted to an university
or an institution of higher education and research.
We extract articles with a title containing the keyword “university” 
in the corresponding language.
Additional extractions with keywords such as \textit{e.g.} "institute", "school", 
"college" have also been performed. 
A manual \textit{a posteriori} check of the automatic 
extraction have been done to remove \textit{e.g} 
fictional universities, colleges and schools of lower education
from the list of top 100 universities of each edition.
We extract also institutions of 
higher education and research which are designated by
acronyms such as \textit{e.g} "ETH Zurich". 
The organizations of pure research (\textit{e.g.}
CNRS, NASA) are not taken into account.

For example in the articles of the 
French Wikipedia edition, ranked by the PageRank algorithm, the first article 
of university  
is entitled ``Université Harvard'' with PageRank index $K=904$, 
then in the second position comes ``École polytechnique (France)'' with $K=1549$, 
and in the third position comes ``Université d'Oxford'' with $K=1558$. 
Thus the top 3 PageRank universities in French Wikipedia 
edition are: 1. Harvard University, 2. École polytechnique, and 3. Oxford University. 
The same procedure is used to rank universities with CheiRank and 2DRank algorithm.
In this way we determine the top 100 universities 
for each of 24 Wikipedia editions. Then each university $U$ obtains 
associated rank index $1 \leq R_{U,E,A} \leq 100$
corresponding to its position in the top 100 list
obtained for edition $E$ by algorithm $A$. 

Following \cite{eomwiki24},
for each type of algorithm, we define a global rank from the rank $ R_{U,E,A}$
of each of 24 Wikipedia editions. Let us define the ranking score \cite{eomwiki24}
\begin{equation}\label{thetascore}
 \Theta_{U,A}=\displaystyle\sum_E \left(101-R_{U,E,A}\right)
\end{equation}
where the summation is done over  24 Wikipedia editions. 
 For a given ranking algorithm, the ranking score 
$\Theta$ of an university will be high if it appears well ranked 
in various Wikipedia language editions.  We use the 
$\Theta$-score (\ref{thetascore}) to merge the 24 world universities 
rankings obtaining the global ranking for all 24 Wikipedia language editions.
We take $R_{U,E,A}=101$ if university is not present in a given edition
thus giving a zero contribution.
The largest value of $\Theta_{U,A}$ determines the first top world university,
the next gives the second world university etc.
Top ranked universities obviously appear in most of the Wikipedia language editions.
For each university we also determine the number of appearances $1 \leq N_a \leq 24$ 
in the top 100 list of universities of each edition.
The global WRWU lists for each algorithm 
are given at the web page \cite{wrwu}.
In total there are $N_u=1025, 1379, 1560$ different universities
for PageRank, CheiRank, 2DRank algorithms respectively.
We notate these global lists as WPRWU, WCRWU, W2RWU respectively.

The top 100 universities for each edition and each 
algorithm are given at \cite{wrwu}. 
To each of these universities we attribute the year of its foundation 
(century), corresponding country (corresponding to
actual country borders given at \cite{worldmap})
and language corresponding to this country
defined in Table~\ref{table2} following the procedure
described in \cite{eomwiki24}. Each university is attributed
to a given country corresponding to the actual borders 
between world countries.
From the global ranks WPRWU, WCRWU, W2RWU we obtain local 
university ranking corresponding to each language
(selection of universities of the same language).
The top 10 universities 
from these  ranks are given
at the web page \cite{wrwu}.
We use the indexes $K_U$ and ${K^*}_U$ for ranks
of global top 100 universities of WPRWU and WCRWU
respectively.

% Results and Discussion can be combined.
\section{WRWU results}

We discuss here the results of the WRWU obtained by the methods
described in the previous Section. The WRWU results are compared
with the ARWU top 100 list taken for year 2013 thus
corresponding to the dating of considered Wikipedia networks.
We also analyze the WRWU in dependence on a country to which a
university is attributed
and consider the geographical distribution
of top universities of WRWU.
The tables of the top 10 universities of
WPRWU, WCRWU, W2RWU, ARWU are presented in 
Tables~\ref{table3},~\ref{table4},~\ref{table5},~\ref{table6}.
These tables have the well known world university.
We discuss the properties of WRWU in more detail in next subsections.

\begin{table}
\caption{
List of the first 10 universities of the Wikipedia 
PageRank of World Universities. The score $\Theta_{PR}$  is defined 
by (\ref{thetascore}); $N_a$ is the number of appearances of a given university 
in the top 100 lists Wikipedia editions.}
\resizebox{\columnwidth}{!}{
\begin{tabular}{llrr}
\hline
Rank&WPRWU&$\Theta_{PR}$&$N_a$\\
\hline
1st	&University of Cambridge	&2272	&24\\
2nd	&University of Oxford	&2247	&24\\
3rd	&Harvard University	&2112	&22\\
4th	&Columbia University	&2025	&23\\
5th	&Princeton University	&1887	&23\\
6th	&Massachusetts Institute of Technology	&1869	&21\\
7th	&University of Chicago	&1783	&22\\
8th	&Stanford University	&1765	&21\\
9th	&Yale University	&1716	&20\\
10th	&University of California, Berkeley	&1557	&19\\
\hline
\end{tabular}
}
\label{table3}
\end{table}

\begin{table}
\caption{
List of the first 10 universities of the Wikipedia 
CheiRank of World Universities, other parameters are as in Table~\ref{table3}. 
}
\resizebox{\columnwidth}{!}{
\begin{tabular}{llrr}
\hline
Rank&WCRWU&$\Theta_{CR}$&$N_a$\\
\hline
1st	&University of Oxford	&1191	&18\\
2nd	&Harvard University	&1025	&17\\
3rd	&Yale University	&1021	&16\\
4th	&Massachusetts Institute of Technology	&816	&16\\
5th	&University of Cambridge	&803	&11\\
6th	&Columbia University	&779	&14\\
7th	&Uppsala University	&751	&11\\
8th	&University of Göttingen	&735	&13\\
9th	&Humboldt University of Berlin	&703	&12\\
10th	&Moscow State University	&699	&14\\
\hline
\end{tabular}
}
\label{table4}
\end{table}

\begin{table}
\caption{
List of the first 10 universities of the Wikipedia 2DRank
of World Universities, other parameters are as in Table~\ref{table3}.
}
\resizebox{\columnwidth}{!}{
\begin{tabular}{llrr}
\hline
Rank&W2RWU&$\Theta_{2R}$&$N_a$\\
\hline
1st	&University of Cambridge	&942	&16\\
2nd	&Columbia University	&786	&11\\
3rd	&Stanford University	&712	&11\\
4th	&Harvard University	&683	&11\\
5th	&Yale University	&609	&11\\
6th	&Princeton University	&596	&11\\
7th	&Massachusetts Institute of Technology	&581	&10\\
8th	&Humboldt University of Berlin	&578	&10\\
9th	&Nanjing University	&516	&8\\
10th	&Johns Hopkins University	&511	&9\\
\hline
\end{tabular}
}
\label{table5}
\end{table}

\begin{table}
\caption{
List of the first 10 universities of ARWU 2013 \cite{shanghai}.
The 3 last columns show the difference between ARWU rank and WPRWU, WCRWU, W2RWU ranks.} 
\resizebox{\columnwidth}{!}{
\begin{tabular}{llrrr}
\hline
Rank&ARWU&WPRWU&WCRWU&W2RWU\\
\hline
1st & Harvard University & -2 & -1 & -3\\
2nd & Stanford University & -6 & -9 & -1\\
3rd & University of California, Berkeley & -7 & -17 & -13\\
4th & Massachusetts Institute of Technology & -2 & 0 & -3\\
5th & University of Cambridge & +4 & 0 & +4\\
6th & California Institute of Technology & -22 & -71 & -124\\
7th & Princeton University & +2 & -15 & +1\\
8th & Columbia University & +4 & +2 & +6\\
9th & University of Chicago & +2 & -45 & -70\\
10th & University of Oxford & +8 & +9 & -2\\
\hline
\end{tabular}
}
\label{table6}
\end{table}

\subsection{Comparison of WRWU and ARWU}

\begin{figure}
\resizebox{\columnwidth}{!}{
\includegraphics{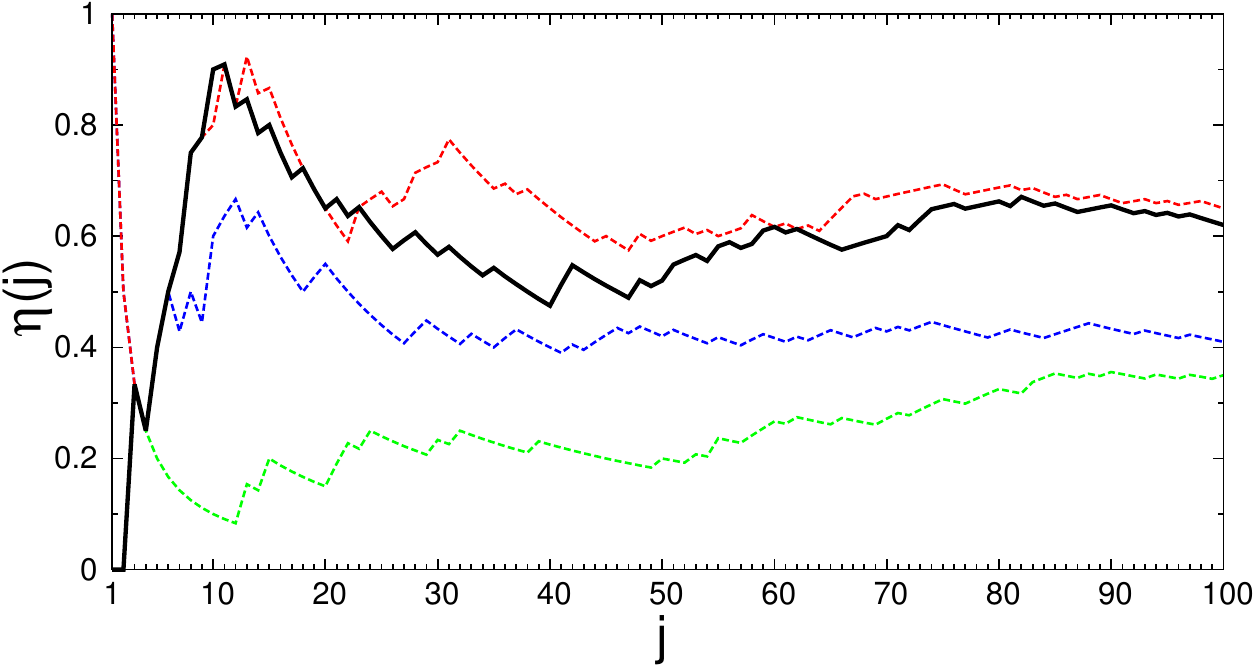}
}
\caption{Overlap $\eta(j)=j_c/j$ of WRWU with ARWU 
as a function of rank index $j$ of ARWU.
Here index  $j_c$ is the number of 
common universities in the top $j$ indexes of  two rankings;
curves show the overlap between: WPRWU and ARWU (black curve),
ARWU and  top PageRank universities of English, French, German (red dashed, blue dashed,
green dashed curves, respectively) Wikipedia editions.
The overlaps with ARWU for WPRWU of English 
and French Wikipedia edition ranks  are superimposed 
up to $j=6$ (black, red and blue dashed curves),
and for ARWU and WPRWU of German  edition the ranks 
are superimposed up to $j=4$ (black and green dashed curves).}
\label{fig1}
\end{figure}
 
\begin{figure}
\resizebox{\columnwidth}{!}{
\includegraphics{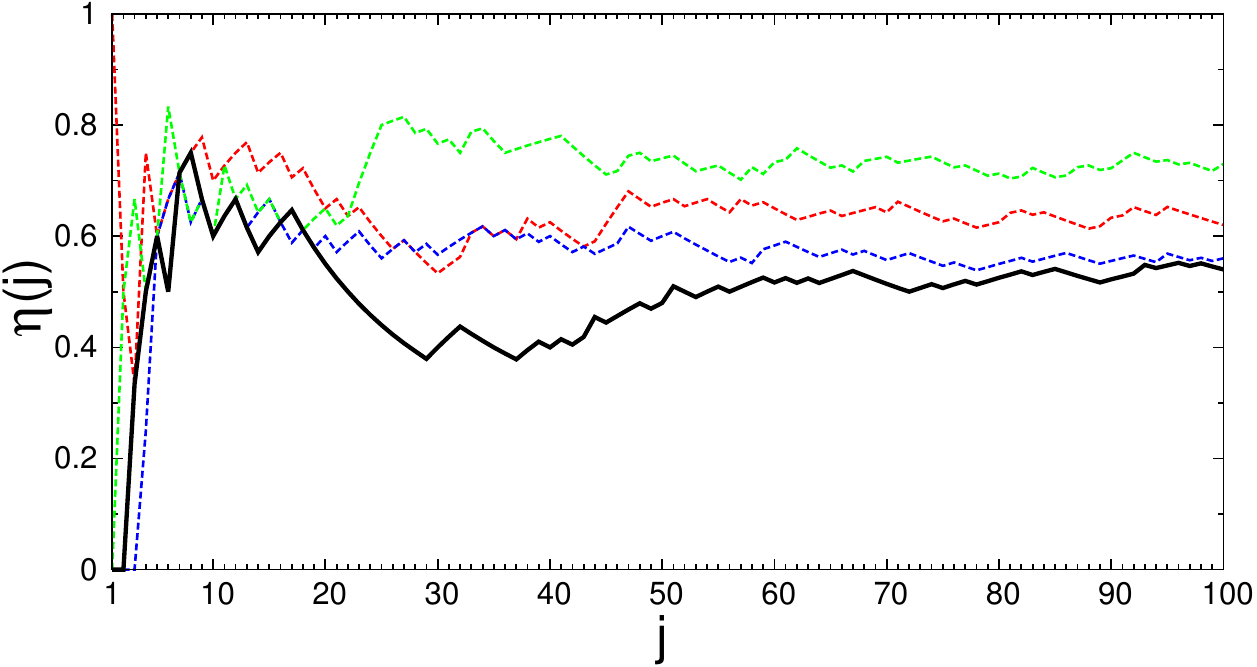}
}
\caption{Overlap  $\eta(j)=j_c/j$ between: W2RWU and
ARWU (black curve),
W2RWU and WPRWU (red dashed curve),
W2RWU and WCRWU (blue dashed curve),
WPRWU and WCRWU (green dashed curve).
Here  $j$ is the rank index of both compared ranks,
$j_c$ is number of common items among $j$ ranks.
}
\label{fig2}
\end{figure}

According to the Tables~\ref{table3},\ref{table4},\ref{table5},\ref{table6}
the overlaps $\eta_{10}$ (fraction of same names among top 10)
with the top 10 list of ARWU are $ \eta_{10} = 0.9, 0.5, 0.6$ 
respectively for WPRWU, WCRWU, W2RWU. Thus WPRWU
gives a reliable ranking of world universities being 
close to the choice of ARWU.
However, at the top positions WPRWU places Cambridge, Oxford and Harvard
which have rank positions $K_{ARWU}=5, 10, 1$ respectively.
We will see later that WRWU gives more favor to ancient
universities, comparing to ARWU. The ranks WCRWU and W2RWU
have  smaller overlap with ARWU. It is related to the fact
that these ranks incorporate the communication features
of the article since CheiRank highlights the effect of outgoing links.
Thus we see that certain university (their articles) 
are not very communicative
(relatively small number of important outgoing links; 
e.g. Chicago, Berkeley)
so that they do not enter in top 10 of WCRWU and W2RWU.
In contrast we see that there are a few non-Anglo-Saxon universities
which gain their higher positions
in W2RWU and WCRWU being more communicative.

The dependence of overlap fractions $\eta(j)$ 
between two university ranks
up to index $j$ is shown in Fig.~\ref{fig1} for  ARWU and WPRWU,
ARWU and PageRank of English, French, and German editions.
For ARWU and WPRWU we find $\eta(100)=0.62$.
It is interesting to note that English edition
has a larger overlap with ARWU ($\eta(100)=0.65$),
followed by French ($\eta(100)=0.41$)
and German ($\eta(100)=0.35$) editions.
Thus we see that ARWU highlights in a stronger way
the contribution of EN universities
while FR and DE editions highlight in a
stronger way the universities of their languages.
Indeed, we have in the top 100 lists
32 French and 63 German speaking universities
for FR and DE editions (see \cite{wrwu}).
This demonstrates significantly different cultural views
developed in each language edition.
We will see below that there are also strong historical 
reasons behind such cultural preferences.

The overlap between ARWU and W2RWU of 2DRank 
is shown in Fig.~\ref{fig2}.
There is a notable reduction of $\eta$ comparing to
the case of WPRWU of Fig.~\ref{fig1}
well visible at $j \approx 10$ and
$j=100$ where we have $\eta(10)= 0.6$, $\eta(100)= 0.54$
instead of larger WPRWU values given above.
The overlaps between W2RWU and WPRWU and WCRWU
lists have larger values
due to certain correlations between 
PageRank and CheiRank vectors
discussed for Wikipedia networks at
\cite{wikizzs,rmpefs,2dmotor}.

\begin{figure}
\resizebox{\columnwidth}{!}{
\includegraphics{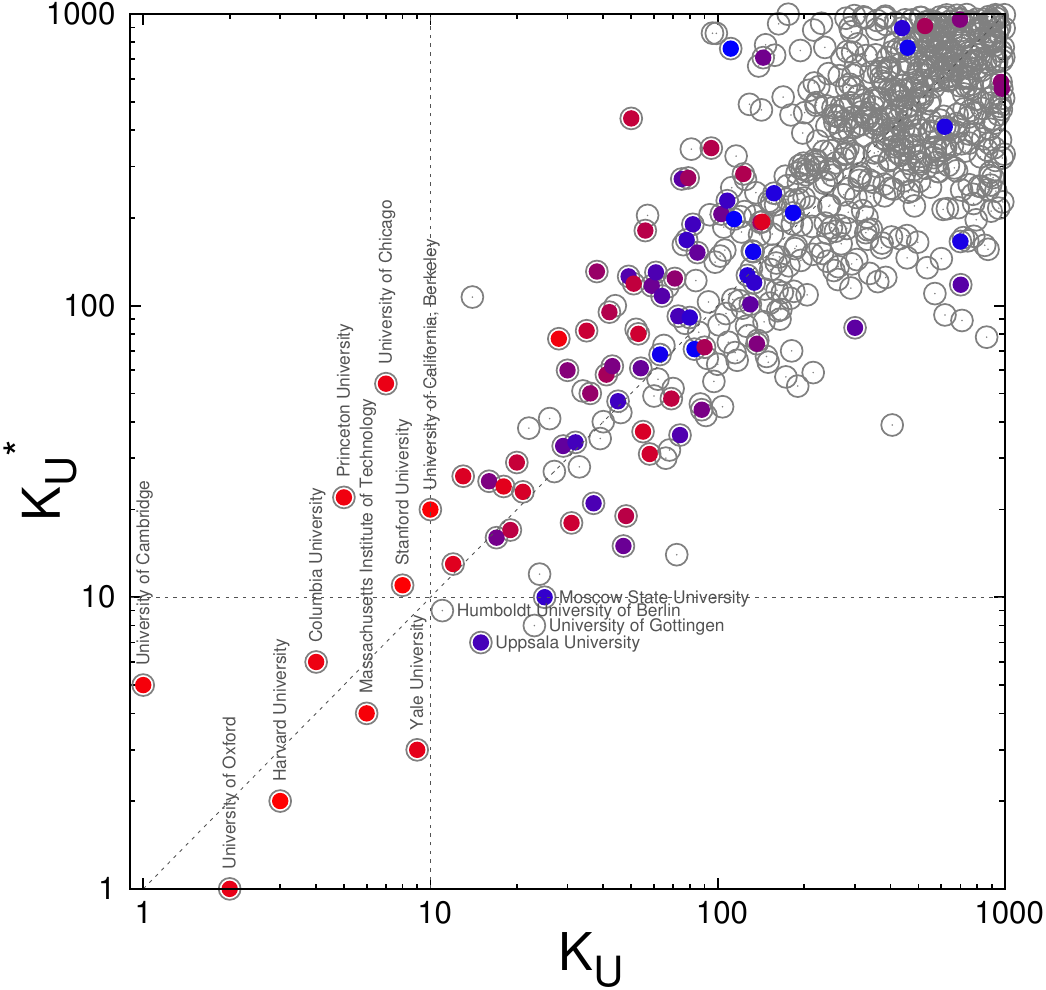}
}
\caption{Distribution of world universities on the PageRank - CheiRank plane
$(K_U,K^*_U)$ (open circles), where $K_U, K^*_U$
are ranks of a given University $U$ in WPRWU, WCRWU. 
Universities appearing in top 100 ARWU Shanghai ranking 
are shown by colored full circles with the color ranging from 
red (ARWU rank 1) to blue (ARWU rank 100).
The names of certain universities are given.}
\label{fig3}
\end{figure}

All 100 universities from 24 editions
are ordered by their respective indexes of WPRWU
PageRank $K_U$ and WCRWU CheiRank  ${K^*_U}$.
Their distribution on the PageRank - CheiRank plane
$(K_U,{K^*_U})$ is shown in Fig.~\ref{fig3}
for the top $1000$ universities.
In contrast to a very broad 
distribution of all Wikipedia articles on this 
plane (see Figs. in \cite{wikizzs,rmpefs,eomwiki9}),
for universities we have significantly stronger
concentration around diagonal
$(K_U={{K^*}_U})$. It looks like that
ingoing and outgoing information
for articles of universities is approximately 
conserved like it is approximately the case of
commercial  flows on the world trade network 
where the countries try to keep
their economic balance \cite{wtn}.
Thus we can say that universities of Oxford, Yale, Uppsala
are more communicative (located below diagonal)
while those of Cambridge, Princeton, Chicago
are much less communicative (located above diagonal).
This presentation shows that certain universities
have open possibilities for improvement of communicative
flows of their Wikipedia articles.

The distribution of ARWU top 100 universities  
is also displayed on $(K_U , K^*_U)$ plane in Fig.~\ref{fig3}. It shows 
that ARWU universities are located mainly at
top $K_U$, $K^*_U$ indexes.
However, some universities,
located at top $K_U$, $K^*_U$ positions, 
are absent in ARWU list. For example,
these are Humboldt and G\"ottingen Universities.
Such cases stress the important difference
between ARWU and WRWU. Namely,
WRWU gives credit to historically 
important universities which played an important role
during the whole human history (e.g. the two cases above
are definitely important for German and EU history)
while ARWU gives much more importance to
instantaneous achievements. We return to discussion
of such differences in next Sections.

Finally, we note that the W2RWU list is not
composed simply from the items of  top squares of 
$(K_U,{K^*_U})$ plane, as in the usual 2DRank algorithm
\cite{rmpefs}, because W2RWU is obtained via the relation 
(\ref{thetascore}) which performs averaging over
24 editions.

\subsection{Geographical distribution}

\begin{figure}
\resizebox{\columnwidth}{!}{
\includegraphics{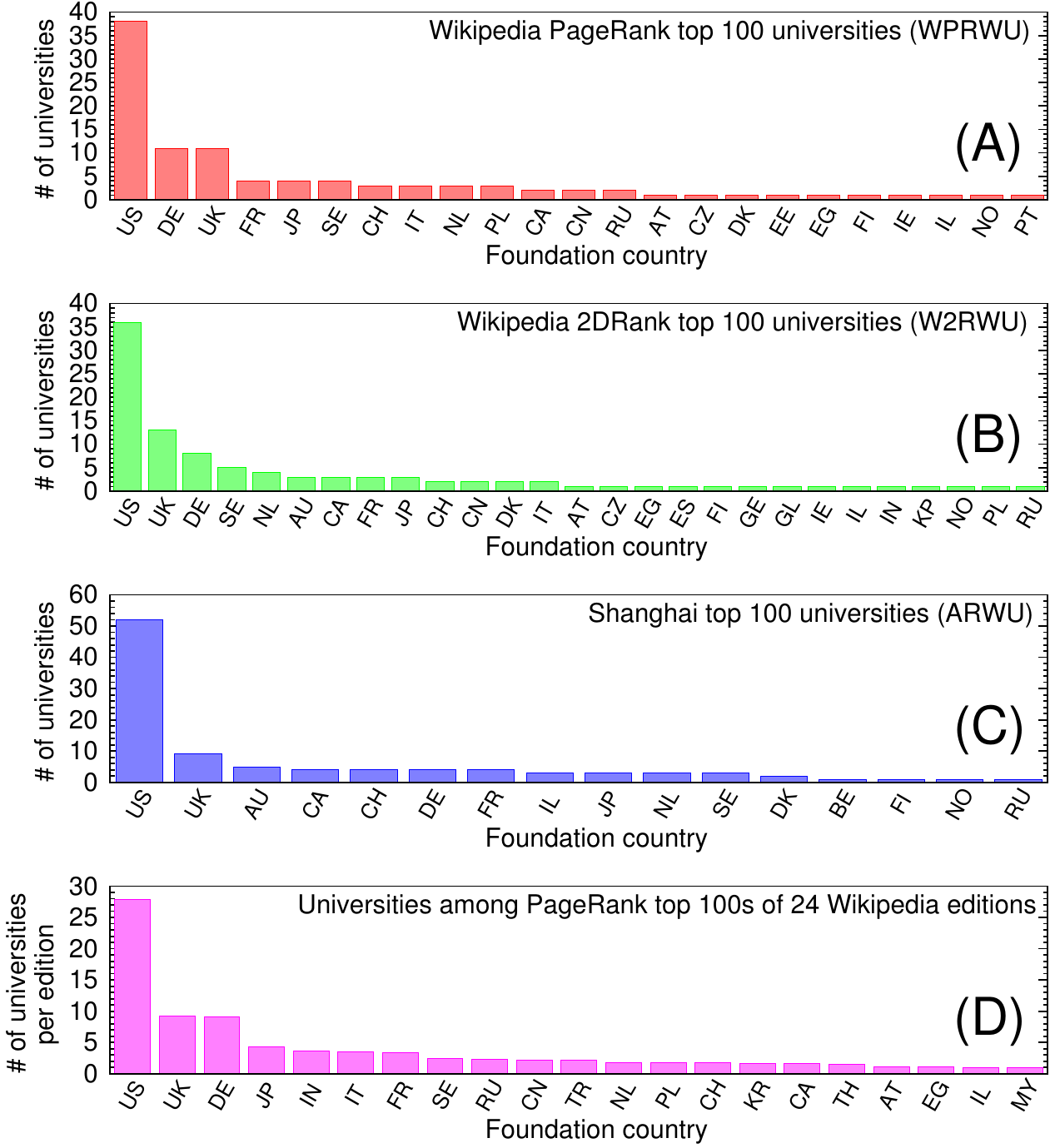}
}
\caption{Distribution over countries of top 100 universities 
from WPRWU \textsf{(A)}, 
from W2RWU \textsf{(B)} and 
from ARWU (Shanghai ranking) \textsf{(C)}.
Panel \textsf{(D)} shows the average (per edition)
number of universities founded 
in a given country appearing in the PageRank top 100 universities of 
24 Wikipedia editions. In panel \textsf{(D)} the
countries with the score less than $1$ are not represented.
In case when countries have equal number of universities they 
are ordered in alphabetic order (see also Section 5).}
\label{fig4}
\end{figure}

According to Table~\ref{table2}
we attribute to each country
a corresponding language edition
choosing mostly used language in a country
defined by actual country borders
(see \cite{worldmap}).
We determine a corresponding country, 
with its actual borders,
at which a university has been founded.
Then from top 100 universities of WPRWU, W2RWU, ARWU 
we obtain their distributions over the world countries
shown in Fig.~\ref{fig4}.
The top 3 countries are US, DE, UK and US, UK, DE
for  WPRWU and  W2RWU
respectively.
This is rather different 
from the top 3 countries US, UK, AU of ARWU.
Also the weight of US  is significantly reduced from 
52  percent for ARWU to
38 and 36  percent for WPRWU and W2RWU
respectively (see Fig.~\ref{fig4}A,B,C).
If we consider the average over all 24 editions
then we get on the top 3 positions
US, UK, DE with even smaller 28.2 percent of US
(see Fig.~\ref{fig4}D). 
Here the rank of country is taken by the namber of its universities in 
the list of top 100. In certain cases
the number of universities happens to be equal
(e.g. for DE and UK for WPRWU),
due to that we use a more refined ranking of countries
as discussed in the las Section.

Thus, in our opinion, Wikipedia ranking
provides a more balanced cultural view on 
important universities. Indeed, each edition
gives more ``votes'' for universities of same
language that increases contributions of 
various languages (or cultures)
even if other cultures do not necessarily
agree on importance of such a choice, thus
introducing certain counterbalance.

%\begin{figure*}
%\resizebox{\textwidth}{!}{
\begin{figure}
\resizebox{\columnwidth}{!}{
\includegraphics{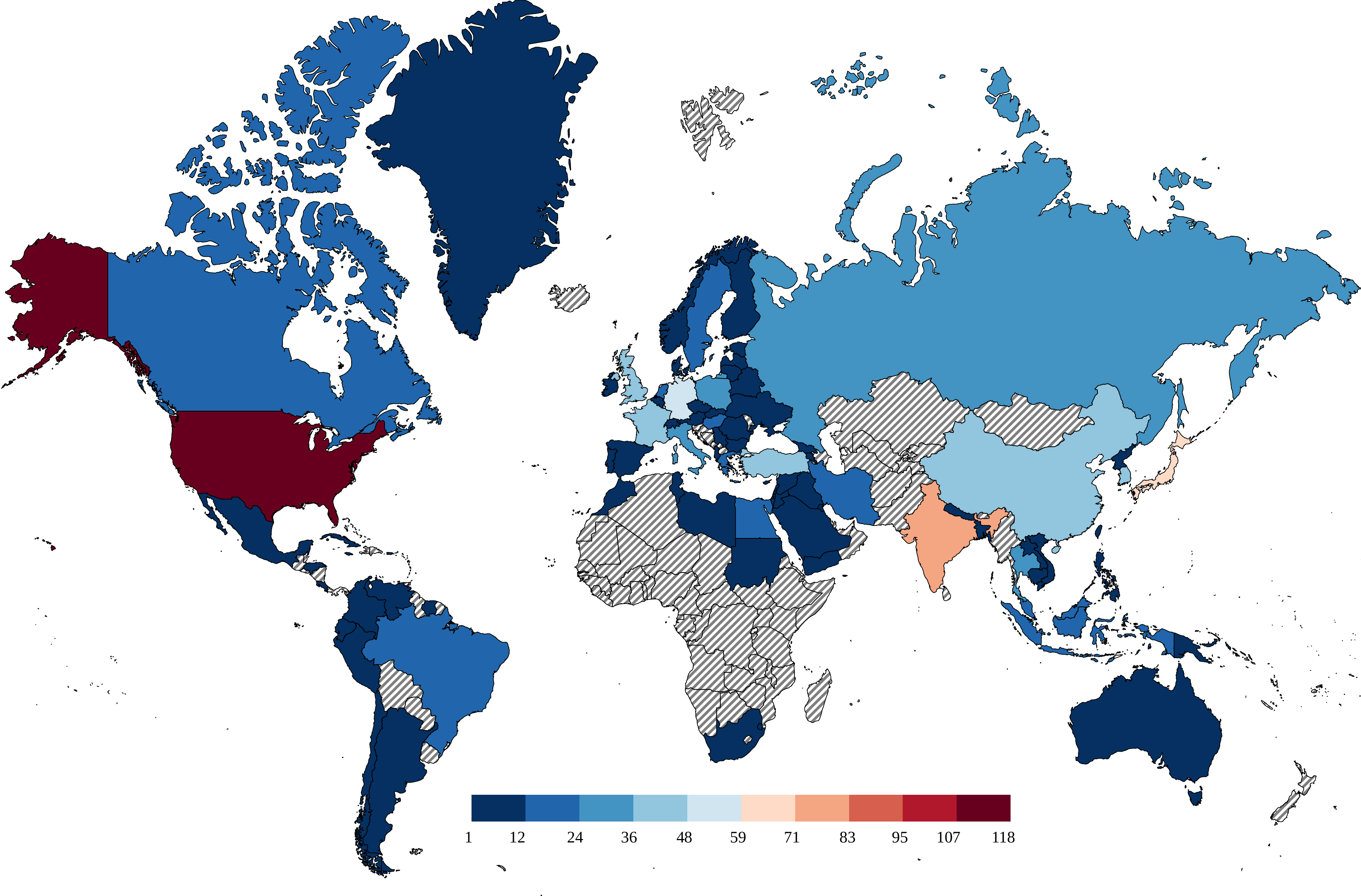}%pdf
}
\caption{Geographical distribution of universities appearing in 
the top 100 universities of all 24 Wikipedia editions 
given by PageRank algorithm.
The total number of universities is $1025$. 
Colors range from dark blue (small number of universities) 
to dark red (maximum number of universities, here $118$ for US). 
Countries filled by dashed lines pattern have 
no university in the top 100 lists of 24 editions.}
\label{fig5}
\end{figure}
%\end{figure*}

\begin{figure}
\resizebox{\columnwidth}{!}{
\includegraphics{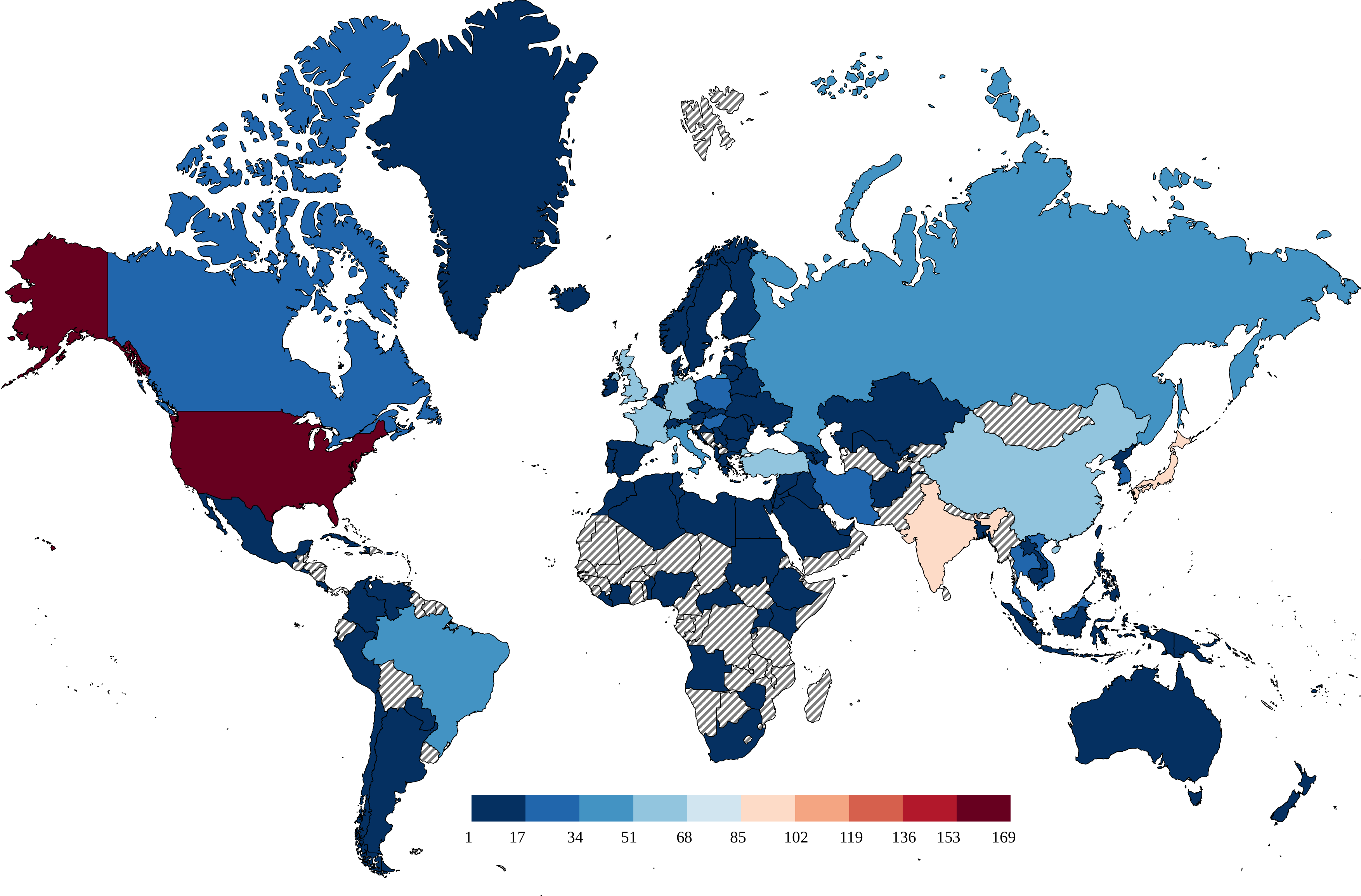}%pdf
}
\caption{Geographical distribution of universities appearing in 
the top 100 universities of all 24 Wikipedia editions 
given by CheiRank algorithm. 
The total number of universities is $1379$.
Colors range from dark blue (small number of universities) 
to dark red (maximum number of universities, here $169$ for US). 
Countries filled by dashed lines pattern have 
no university in the top 100 lists of 24 editions.}
\label{fig6}
\end{figure}

\begin{figure}
\resizebox{\columnwidth}{!}{
\includegraphics{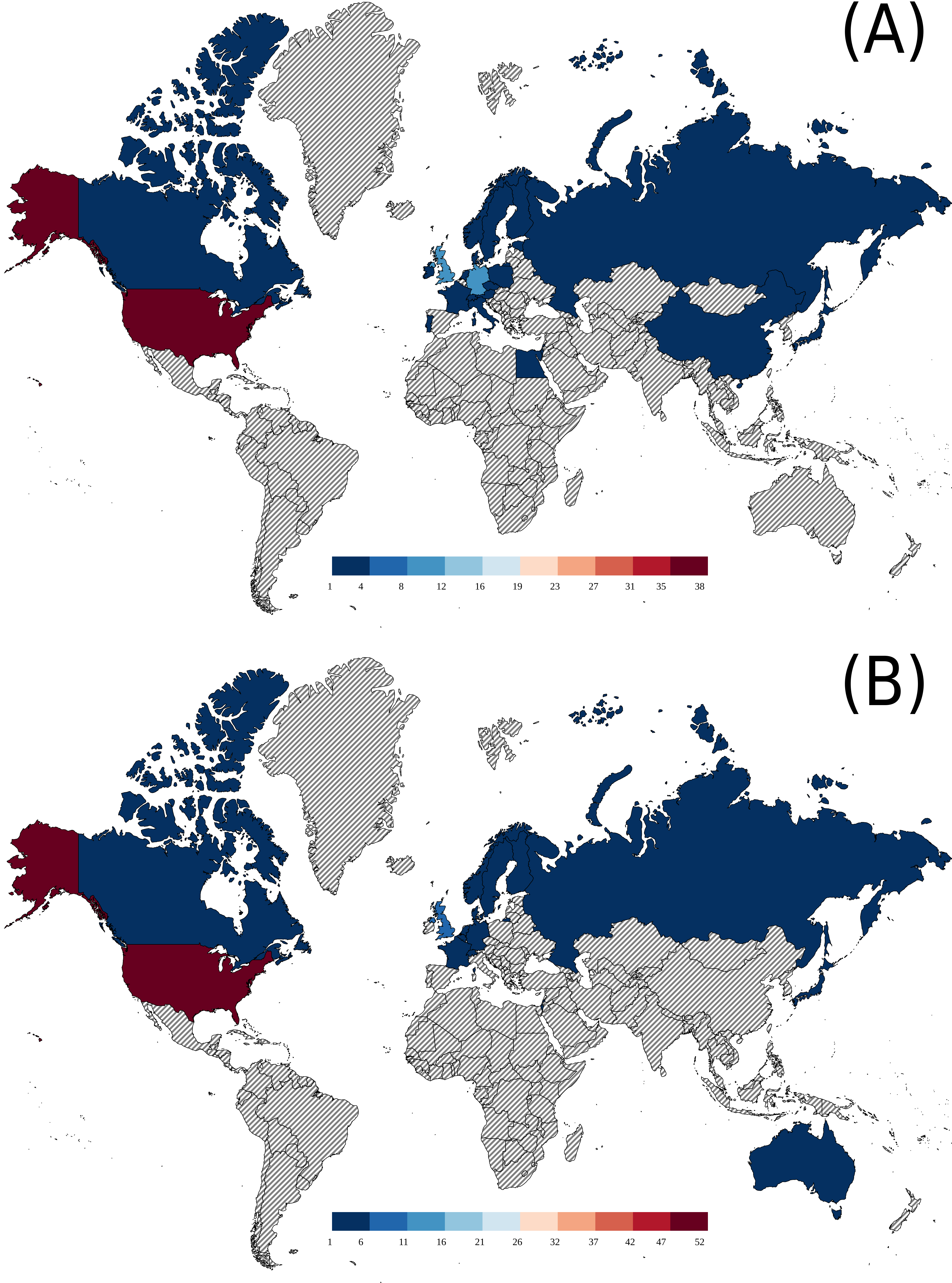}%pdf
}
\caption{Geographical distribution of top 100 universities 
of WPRWU from PageRank algorithm (A)
and of ARWU Shanghai ranking (B).
Colors range from dark blue 
(small number of universities) to 
dark red (maximum 38 (A) and 52 (B) for US). 
Countries filled by dashed lines pattern 
have no university in corresponding top 100.}
\label{fig7}
\end{figure}

The distribution of all top 100 PageRank universities of 24 editions
over the world map of countries \cite{worldmap}
is shown in Fig.~\ref{fig5}. 
The similar world map for WCRWU is shown in Fig.~\ref{fig6}.
For WCRWU we see appearance of new countries in Africa
and Central Asia being related to a larger number of 
outgoing links of their universities.
Differently from top 100 of WPRWU (or WCRWU),
here all 24 editions give a more significant contribution
with a noticeable weight for India and Japan. 
Indeed, in HI and JA edition rankings there are
large fractions of universities of their own languages
(81 and 65 percent respectively) that leads to their
weight increase among all 1025 PageRank universities.
However, this effect of self citations
is significantly suppressed for top 100 of WPRWU
where opinions of other editions play a role.
Thus Indian universities do not appear in the WPRWU top
100 as it is seen in Fig.~\ref{fig4} 
and in the corresponding world map of Fig.~\ref{fig7}(A)
showing WPRWU top 100 universities.

The geographical distributions of WPRWU and ARWU
are shown in Fig.~\ref{fig7} (A) and (B). 
We see that Australia, present at high positions in ARWU,
is not present at WPRWU, while inversely
China is present on WPRWU map being absent  in ARWU.
Also an absolute percent of US is significantly 
reduced in WPRWU with Germany
taking the second position at WPRWU
instead of 6th position in ARWU.

\begin{figure}
\resizebox{\columnwidth}{!}{
\includegraphics{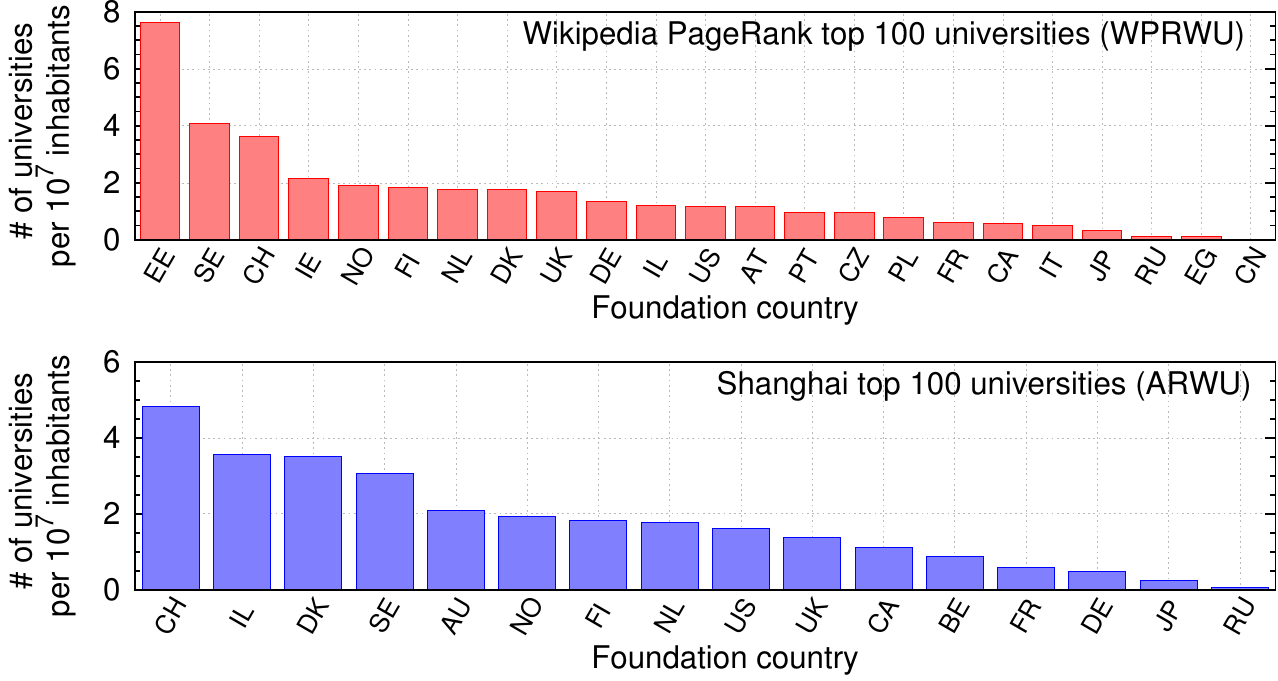}
}
\caption{Distributions  of WPRWU and ARWU top 
100 universities per number of inhabitants
(measured in $10$ millions)
over corresponding country.}
\label{fig8}
\end{figure}

Another interesting comparison of efficiency
of universities is given by a number of top 100
universities per 10 millions inhabitants for 
a given country (the actual country population is 
taken mainly in 2015 from \cite{wikipopul}, see also \cite{wrwu}). 
These distributions
for highly ranked countries are shown in Fig.~\ref{fig8}
for WPRWU and ARWU.  At the top 3 positions
we find Estonia, Sweden, Switzerland for WPRWU
and Switzerland, Israel, Denmark for ARWU.
Estonia appears on the top due to its small population
and the only one University of Tartu. This ancient university, 
founded in 1632, was historically on the
crossroads of Sweden, Russia, Poland, Germany
thus being important for various cultures in this
region. Now it is located in Estonia with its
small population that explains its top WPRWU per inhabitant
position. The example of University of Tartu highlights the importance of
historical environment for appreciation of role of a given university.
It shows that WPRWU takes into account the history of university
while ARWU ignores this feature.

On the second WPRWU position per inhabitant we have Sweden
with 4 universities in top 100. 
Sweden is followed by Switzerland with 3 universities.
For ARWU Switzerland is at the top position
with 4 universities. We see  that the 
Northern countries are taking high positions
(SE, IE, NO, FI ...) 
both for WPRWU and ARWU.

We now go to analysis of time evolution of
top universities.

\subsection{Evolution through centuries}

\begin{figure}
\resizebox{\columnwidth}{!}{
\includegraphics{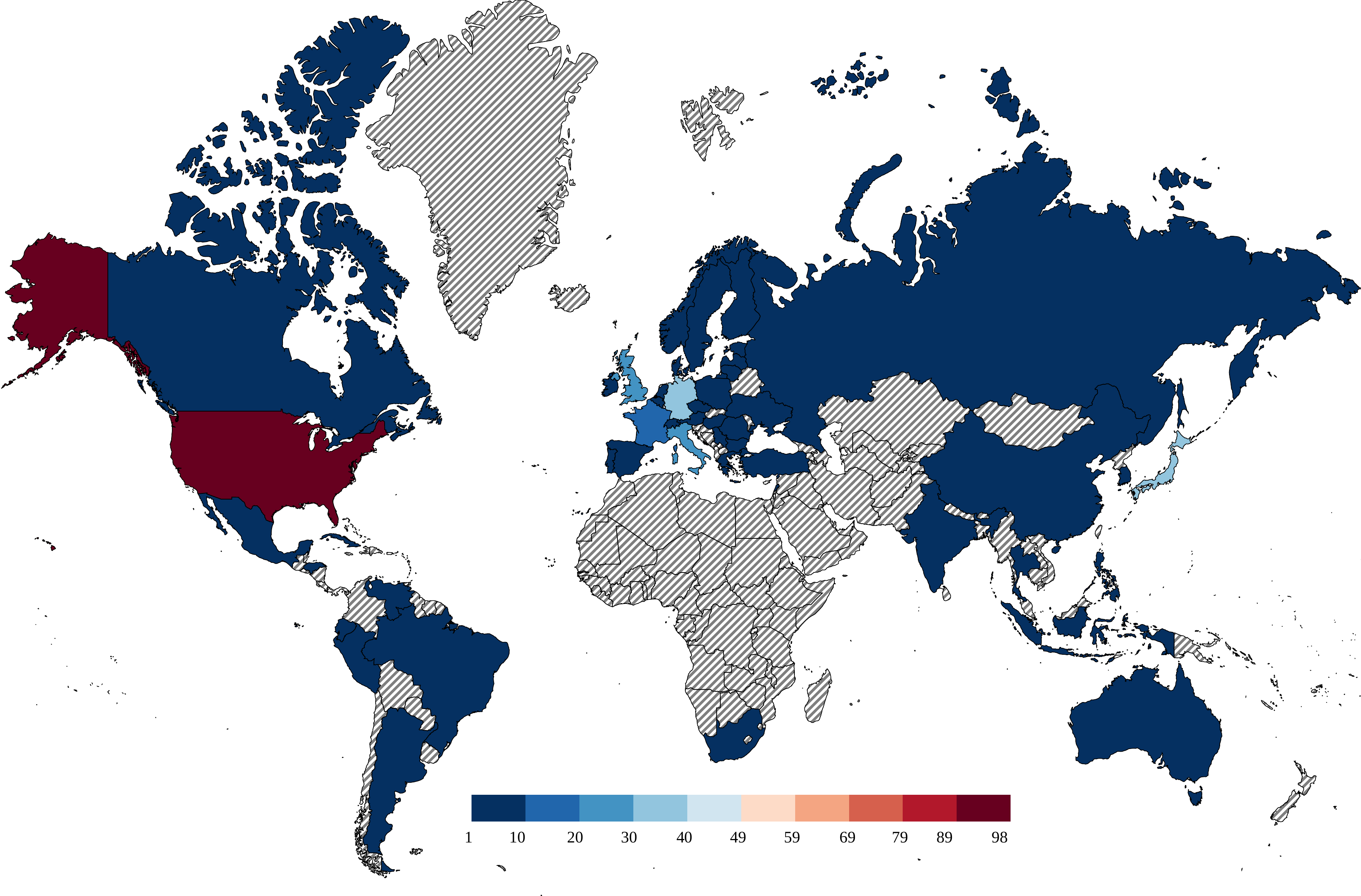}%pdf
}
\caption{Geographical distribution of universities appearing in 
the top 100 universities of all 24 Wikipedia editions 
given by PageRank algorithm and founded before 20th century
(384 in total).
Colors range from dark blue (small number of universities) 
to dark red (maximum number of universities, here $98$ for US). 
Countries filled by dashed lines pattern have 
no university in the top 100 lists of 24 editions.}
\label{fig9}
\end{figure}

\begin{figure}
\resizebox{\columnwidth}{!}{
\includegraphics{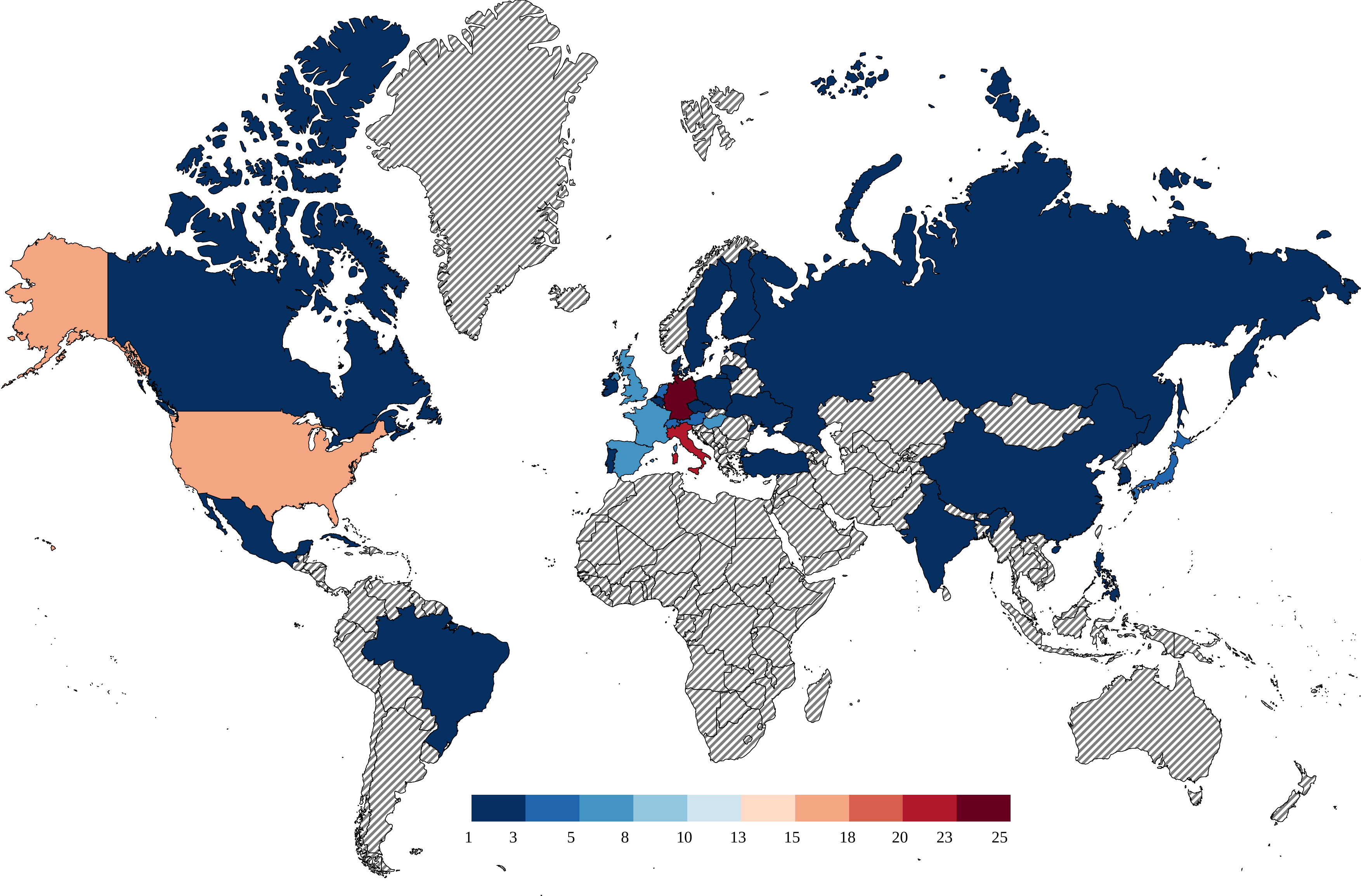}%pdf
}
\caption{Same as in Fig.~\ref{fig9} but with the foundation date before 19th century
(139 universities in total; maximum is for DE with 25 universities).}
\label{fig10}
\end{figure}

Each university has its year of foundation
and thus we attribute all universities to their
own foundation century. From top 100 universities of
all editions with PageRank we select universities founded before 20th
century (384 in total) and present their geographical distribution 
over world countries in Fig.~\ref{fig9}. The comparison with Fig.~\ref{fig5}
shows a significant drop of number of universities in Africa 
(only South Africa remains), India and Japan lose their high positions
while EU countries (DE, UK, IT, FR) are improving their positions
but US still takes the first top position
with the largest number of universities. 
The geographical distribution for universities founded before 19th century
(139 in total) is shown in Fig.~\ref{fig10}.
Here Germany is taking the top position followed by Italy.
But already US takes the 3rd position.

\begin{figure}
\resizebox{\columnwidth}{!}{
\includegraphics{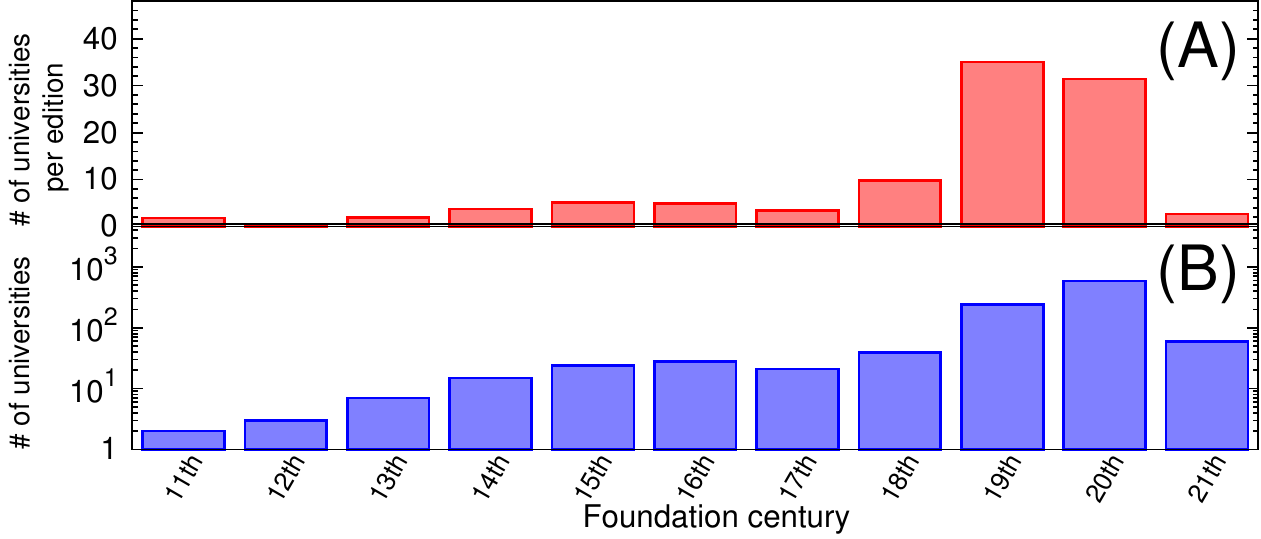}
}
\caption{Distribution over foundation century
for universities $N_f$,
appearing  in the PageRank top 100 universities of 
24 Wikipedia editions (1025 in total);
panel (A) gives the average number per edition $N_{fe}$,
panel (B) gives the total number of universities $N_f$
founded in a given century.
}
\label{fig11}
\end{figure}

\begin{figure}
\resizebox{\columnwidth}{!}{
\includegraphics{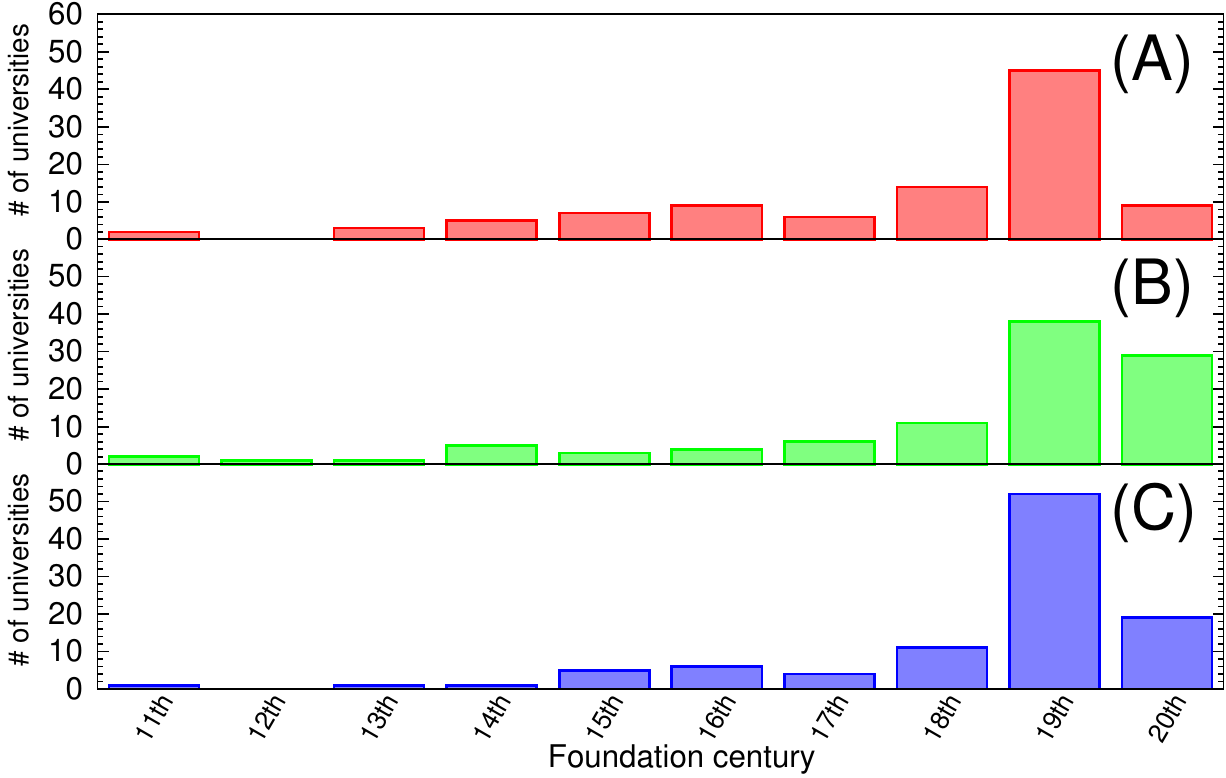}
}
\caption{Distribution over foundation century
of top 100 universities from WPRWU (A),
W2RWU (B) and ARWU (C).}
\label{fig12}
\end{figure}

The distribution of number of universities $N_f$ founded
in a given century is given in Fig.~\ref{fig11}.
It shows  that both, the total number $N_f$ and 
average number of universities per edition $N_{fe}$,
remain approximately constant during 14th to 17th centuries.
The steady growth of $N_f$  starts from 18th century
being close to an exponential increase.
On the other side if we consider only top 100
of global ranking of WPRWU, W2RWU and ARWU,
shown in Fig.~\ref{fig12},
then we see that the main part of top universities 
has been founded in 19th century (about 50)
and in 20th century there appeared only about 10 to 20 
universities which succeed to enter in the top 100 list.
Thus we see that the top 100 club is rather rigid
in accepting new ``members'' with time.
Only for W2RWU there is some redistribution in 20th century
mainly because the new young universities 
are more communicative that improves their 2DRank positions.
In contrast for WPRWU the first 43 (5) universities 
have been founded before
20th (19th) century, and which remained 
de facto at unchanged positions till now.
In total in WPRWU top 100 there are 9, 45, 14, 6 universities founded
in 20th, 19th, 18th, 17th centuries.
This confirms the highly rigid nature of top
100 positing of leading universities
which was mainly formed before 20th century.

\begin{figure}
\resizebox{\columnwidth}{!}{
\includegraphics{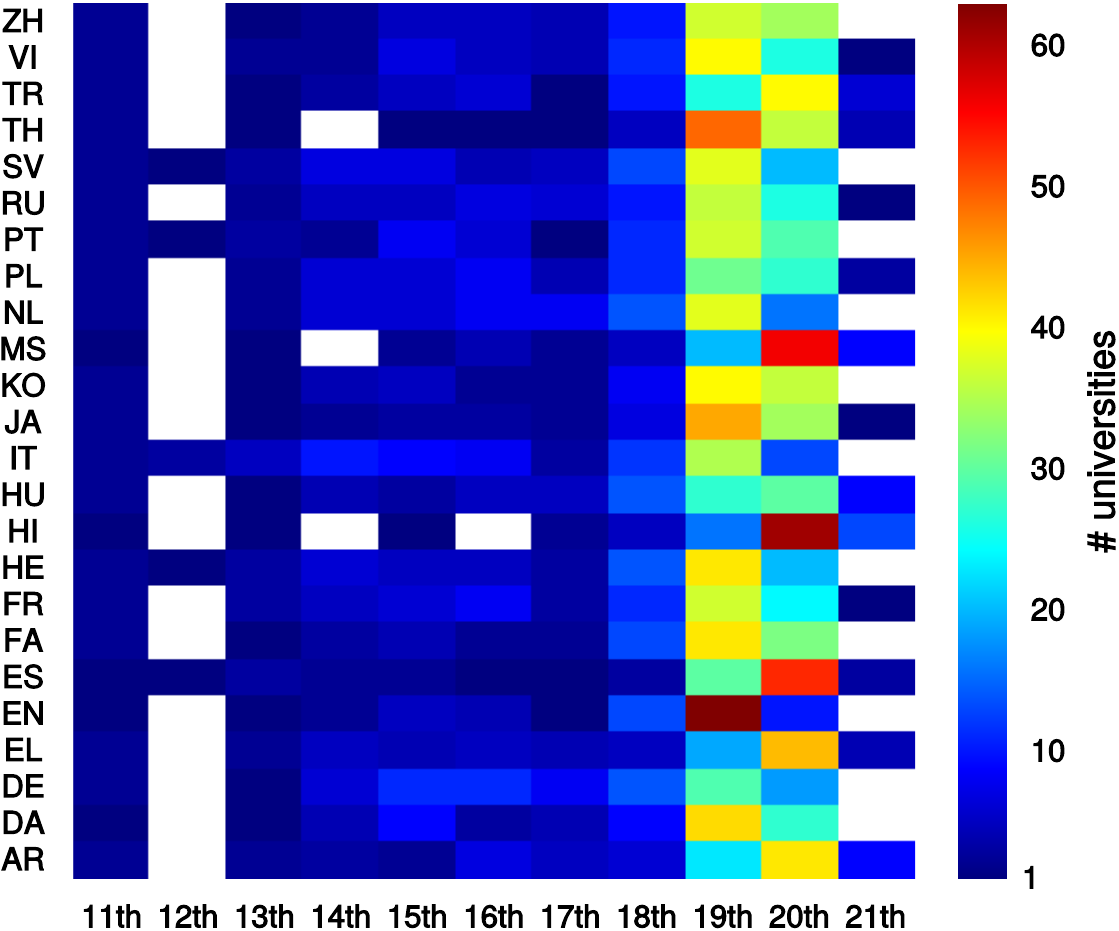}
}
\caption{Distribution over foundation century
for PageRank top 100 universities of 
each Wikipedia edition. 
Dark red color corresponds to maximum 
with 63 for EN in 19th century, dark blue color corresponds to one university,
white color represents zero.}
\label{fig13}
\end{figure}

\begin{figure}
\resizebox{\columnwidth}{!}{
\includegraphics{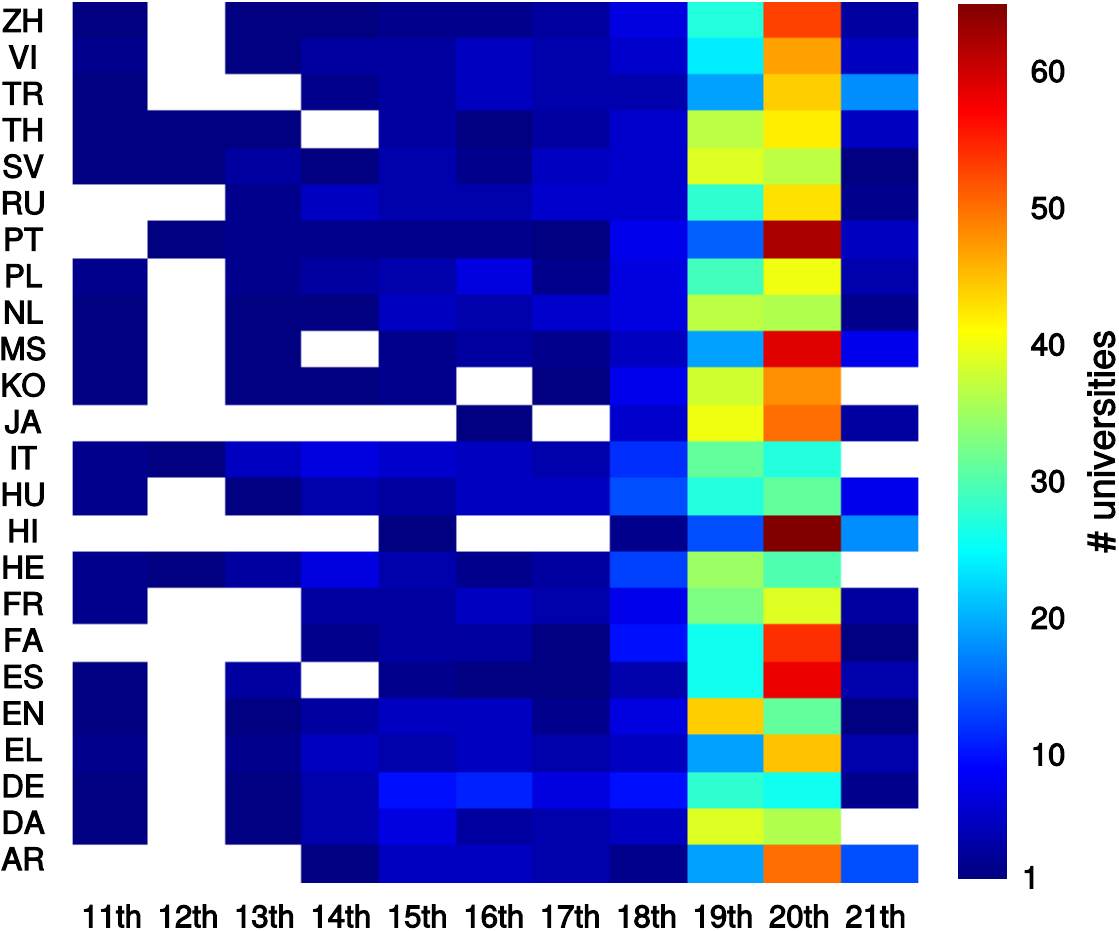}
}
\caption{Distribution over foundation century
for CheiRank top 100 universities of 
each Wikipedia edition. 
Dark red color corresponds to maximum 
with 65 for HI in 20th century,
dark blue color corresponds to one university,
white color represents zero.}
\label{fig14}
\end{figure}

In Fig.~\ref{fig12} we consider the global list of top 100
universities. The data for top 100 of each edition at each 
foundation century are presented in Fig.~\ref{fig13}
for PageRank and in Fig.~\ref{fig14} for CheiRank.
The data of Fig.~\ref{fig13} show emergence of many 
universities in PageRank top 100
in 20th century for HI, MS and ES
(only 3 editions with more 
then 50 universities in this century). 
In contrast,
there are practically no new universities 
in 20th century for EN, IT, NL showing that 
their contribution to the top 100 list
is dominated by 19th century. 
The situation is different for the CheiRank top 100 in Fig.~\ref{fig14}:
here there are many universities appearing in 20th century
especially for HI, PT, MS and even ES etc (8 editions
with more then 50 universities in this century).
We attribute this to the fact that CheiRank
highlights the communicative features of
Wikipedia articles and that the new young universities 
are more better placed in communicative
broadcast activity.

Finally, for an interested reader we give the names of 
early founded universities: U Oxford, U Bologna (century XI);
U Salamanca, U Modena and Teggio Emilia, U Parma (century XII);
U Cambridge, U Padua, U Coimbra, U Naples Federico II,
Complutense U Madrid, U Siena, U Lleida (century XIII).

\section{Entanglement and interactions of cultures}

The results for Wikipedia ranking in different editions can be used 
for analysis of entanglement and interactions of cultures.
We associate each language to a culture since it represents 
the most important cultural feature.
Such an approach was used in \cite{eomwiki9,eomwiki24}
for historical figures and this method can be also directly used for 
universities. For that we count how many 
universities $N_{ij}$, attributed to a culture (language) $i$, 
appears in the top 100 universities of culture (language) $j$.
This gives the number of directed links from node $j$
to node $i$ and then the Google matrix of cultures is constructed
in the usual way (\ref{eq:gmatrix}) with the same 
damping factor $\alpha=0.85$ (see also \cite{eomwiki9,eomwiki24}).
In total we have 25 nodes,  since some universities
cannot be attributed to any of 24 editions and 
corresponding cultures and in such a case 
we attribute them to an additional culture WR.
The diagonal self citations inside the same culture are
not considered so that $N_{ii}=0$, $1 \leq i \leq 25$. 
The number $N_{ij}$ can be defined for any of the 
three ranking algorithms discussed above.

\begin{figure}
\resizebox{\columnwidth}{!}{
\includegraphics{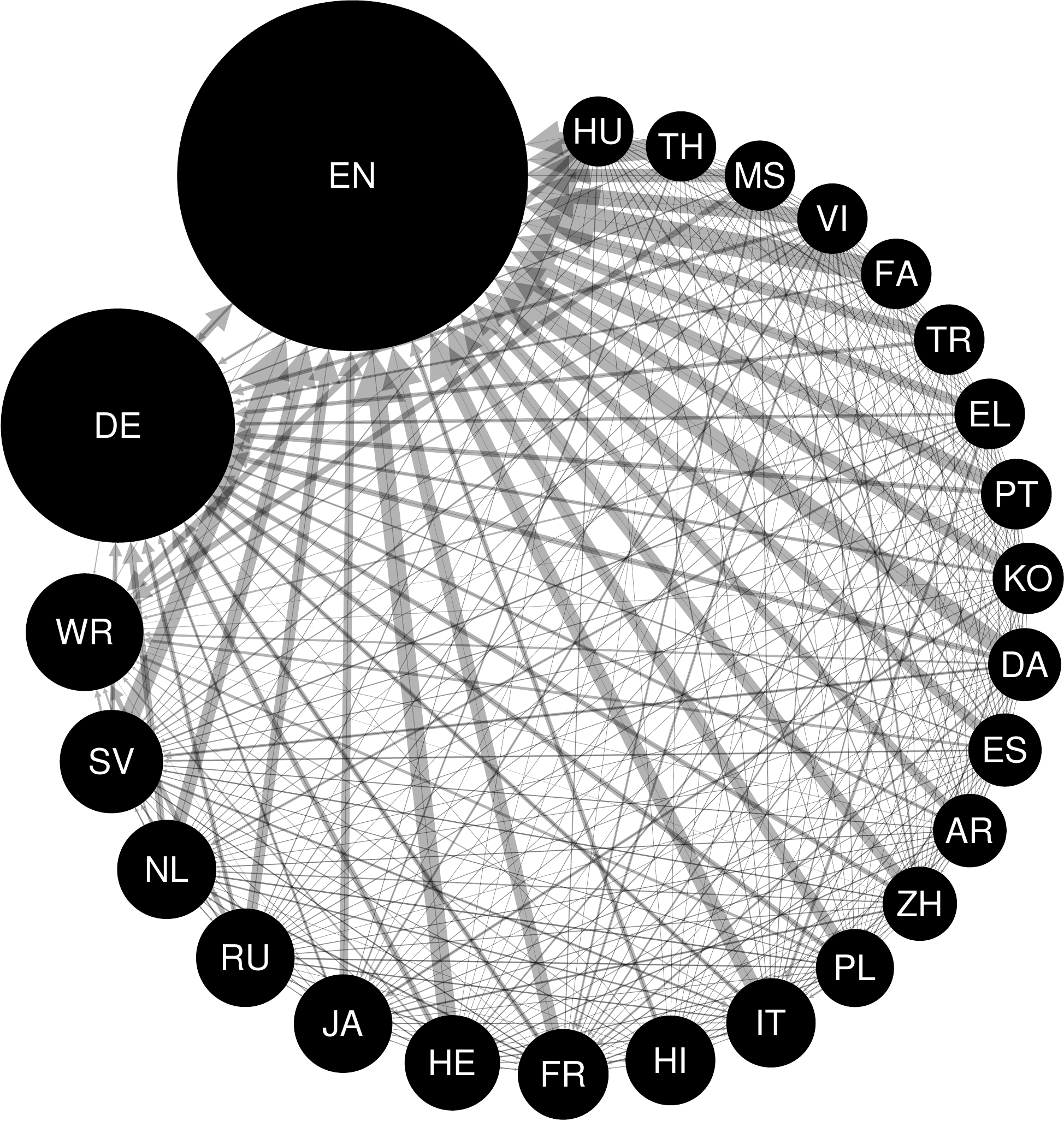}
}
\caption{Network of cultures constructed from the PageRank top 100 universities 
of 24 Wikipedia editions.
The width of each directed link is proportional to the number of 
foreign universities quoted in top 100 of a given culture;
%a link 
%direction is clockwise and 
%goes from a given culture to a culture 
%of quoted foreign universities,
links inside cultures are not considered. 
The size of a node is proportional to its PageRank value.
Two letters in the node encode the language used
with the exception of the node 'WR'  which represents all
other languages besides the 24 languages represented in the graph.}
\label{fig15}
\end{figure}

\begin{figure}
\resizebox{\columnwidth}{!}{
\includegraphics{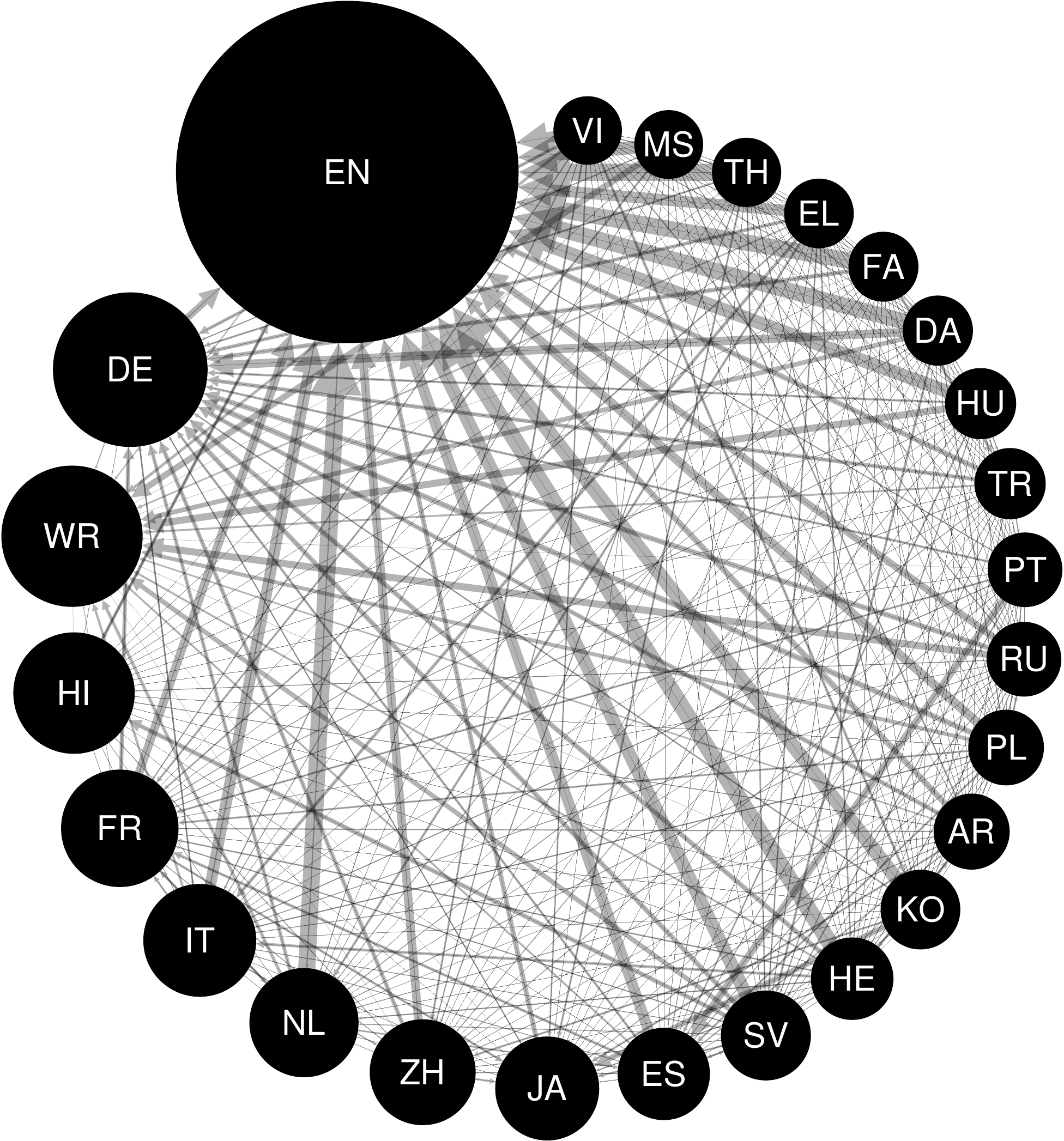}
}
\caption{Same as in Fig.~\ref{fig15} but for the CheiRank top 100 universities
of 24 Wikipedia editions.}
\label{fig16}
\end{figure}

The network of cultures, constructed from
top 100 universities of PageRank,
is shown in Fig.~\ref{fig15}. For the CheiRank algorithm
the network of cultures is shown in Fig.~\ref{fig16}.
For the presentation of these directed networks we use gephy software
with a circular layout
%the Fruchterman–Reingold drawing algorithm
\cite{networkimage}.
These network images show strong interconnections 
between different cultures. 

\begin{figure}
\resizebox{\columnwidth}{!}{
\includegraphics{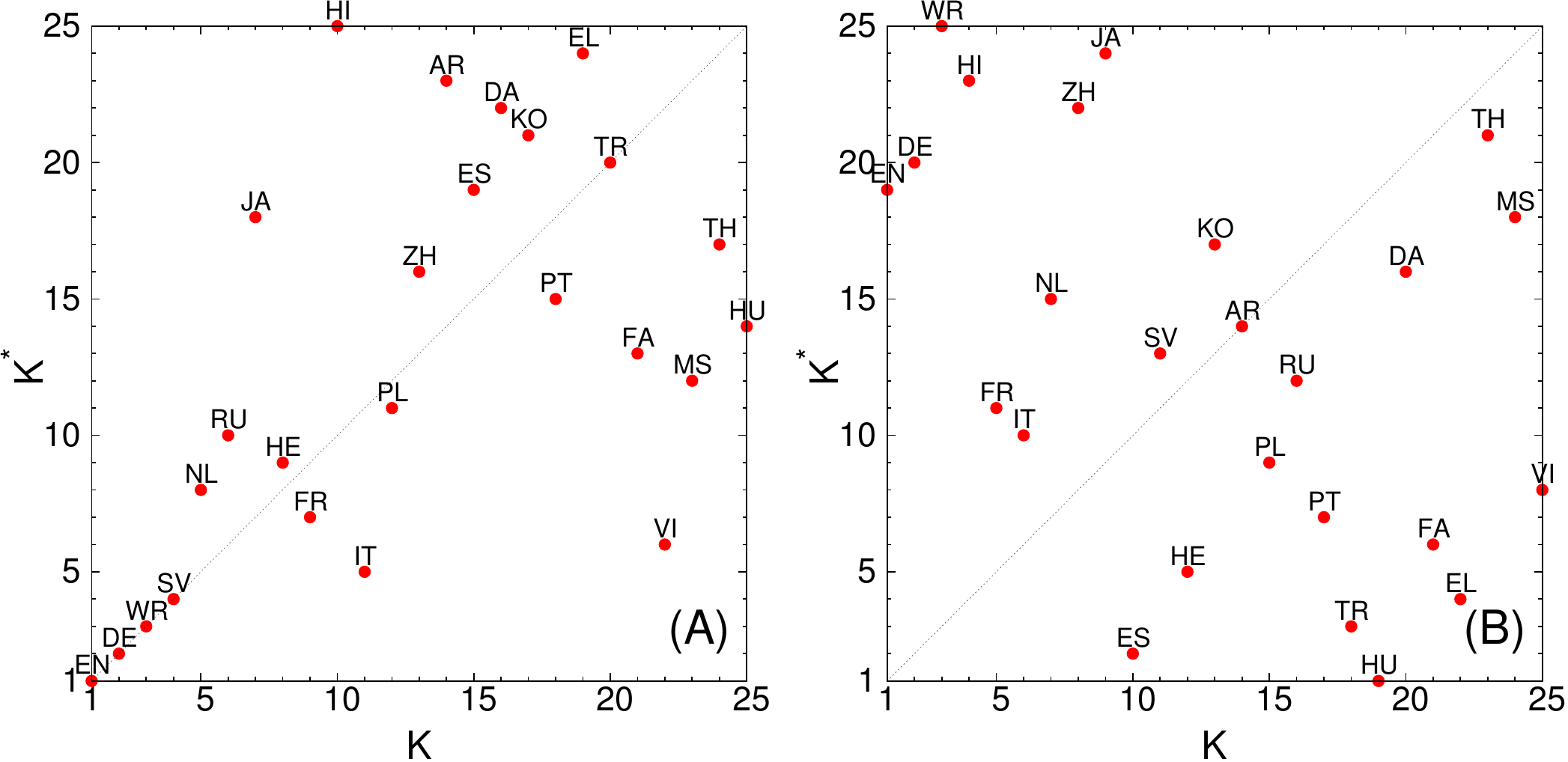}
}
\caption{PageRank-CheiRank $(K,K^*)$ plane
for the networks of cultures from PageRank universities
of Fig.~\ref{fig15} (panel A) and 
from CheiRank universities of Fig.~\ref{fig16} (panel B).
Each culture is marked by language code.
}
\label{fig17}
\end{figure}

To analyze this complex entanglement of cultures
we determine the PageRank and CheiRank vectors
of such a secondary network with 25 nodes.
The size of a node in Fig.~\ref{fig15} (Fig.~\ref{fig16})
is proportional to PageRank (CheiRank) probability.
In addition we determine  PageRank $K$ and CheiRank $K^*$
indexes of cultures and display all $24+1$ cultures on the
$(K,K^*)$ plane in Fig.~\ref{fig17}.
For PageRank list (Fig.~\ref{fig17}A) we have the strong dominance
of EN, DE, WR, and SV cultures which take 
the top 4 positions being on the diagonal 
$K=K^*$. Next positions in $K$ are taken by NL, RU
and in $K^*$ by IT, FR. At highest positions
of $K$ and $K^*$ we have HI and HU
which have many self citations.
For CheiRank list  (Fig.~\ref{fig17}B)
we have significantly broader 
distributions of cultures on $(K,K^*)$ plane
with EN, DE in top $K$ positions and
HU, ES in top $K^*$ positions.
We assume that many well-know scientists immigrated
from Hungary, and many Spanish speaking countries in Latin America
are responsible for stronger communicative features of
HU and ES while
EN and DE still keep their top PageRank positions.
We note that for ranking of universities we
obtain the distributions of cultures over
$K,K^*$ plane being rather different from the
distributions of cultures obtained from ranking of historical figures
(see Fig.10 in \cite{eomwiki24}). This shows that 
the entanglement of cultures takes place on various
levels of knowledge having complex 
interactions on each level.
Thus appreciation of foreign universities in a given culture
works in a rather different manner comparing
to appreciation of foreign historical figures.

\section{Discussion}

In this work we presented the Wikipedia ranking of world
universities using PageRank, 2DRank and CheiRank algorithms
developed for directed networks where they proved their
efficiency. The analysis is based on 24 Wikipedia language editions
that allows to take into account various cultural view points.
At the same time these cultural views are 
considered by the statistical mathematical analysis of all human
knowledge accumulated in these 24 editions containing 17 million
Wikipedia articles. Thus our analysis gives no cultural preferences
standing on pure mathematical grounds.

We find that the PageRank list of WPRWU top 100 universities
has 62 percent overlap with ARWU Shanghai list
demonstrating that this analysis gives reliable results.
At the same time WPRWU gives more emphasis to non-Anglo-Saxon
cultures reducing the percent of US universities from 52 in ARWU
to 38 in WPRWU. The ranking of world countries
is obtained by counting a country score 
by summation of scores of universities belonging
to a given country. The top 10 countries are given in 
Table \ref{table7}.
Our results show that UK  
takes the 2nd position in this ranking with 
Germany, Sweden, France, Japan taking next 3rd to 6th places,
while ARWU gives respectively the places
7th for DE, 2nd for UK, 6th for FR, 5th for JP
and 9th for SE. The number of top PageRank universities per 
inhabitant demonstrates the  efficiency of universities in 
countries of Northern Europe and Switzerland.

\begin{table}
\caption{Ranking of countries for universities in top 100 WPRWU and in 
top 100 ARWU.
Only first ten countries are shown, the complete lists is available at 
\cite{wrwu}.
Ranking of countries is established using the country score 
$\Theta_{C,R}=\sum_{U}(101-K_{U,C,R})$
where the sum is over universities of a given country $C$ and where 
$K_{U,C,R}$ is the rank of university $U$
in ranking $R$; left (right) columns correspond to $R$=WPRWU ($R$=ARWU).
}
\resizebox{\columnwidth}{!}{
\begin{tabular}{llrr|llrr}
\hline
\multicolumn{4}{c|}{WPRWU}&\multicolumn{4}{c}{ARWU}\\
\hline
Rank&CC&$\Theta_{C,WPRWU}$&$N_C$&Rank&CC&$\Theta_{C,ARWU}$&$N_C$\\
\hline
1st    &US    &2098    &38    &1st    &US &3100    &52\\
2nd    &UK    &640    &10  &2nd    &UK &531    &9\\
3rd    &DE    &566    &12  &3rd    &CA &182    &4\\
4th    &SE    &205    &4  &4th    &CH &168    &4\\
5th    &FR    &159    &4  &5th    &JP &166    &3\\
  6th    &JP    &155    &4  &6th    &FR &156    &4\\
  7th    &IT    &155    &3  &7th    &DE &133    &4\\
  8th    &CH    &122    &3  &8th    &AU &101    &5\\
  9th    &RU    &110    &2  &9th    &SE &101    &3\\
  10th    &CA    &95    &2  &10th    &NL &78    &3\\
\hline
\end{tabular}
}
\label{table7}
\end{table}

The rankings based on 2DRank and CheiRank algorithms
highlight in a better manner the communicative and broadcast features
of universities showing that their efficiency varies strongly even for 
top ranked universities. 

The analysis of university ranking evolution through ten centuries shows
that  Wikipedia highlights significantly  stronger historically 
important universities  which role is reduced in ARWU.
Thus for PageRank list of top 100 universities in 24 editions
we find the dominance of Germany and Italy before 19th century,
even if the rise of US universities is already visible to that times.
The dominance of US is established after 19th century.
Our WRWU results show that the club of top 
universities is formed mainly before 20th century
and that it remains very ridig in
``accepting'' new members after that time.

The appreciation of foreign universities in individual editions
allows to determine effective interactions of 24 cultures
related to language editions showing the strong influence
of English, German and Swedish universities.

We think that the Wikipedia ranking 
provides the firm mathematical statistical 
evaluation of world universities which
can be viewed as a new independent 
ranking being complementary to already
existing approaches.
In the view of importance of
university ranking for higher education \cite{hazelkorn}
we hope that the WRWU method will also find
a broad usage together with other rankings.

% \begin{figure}
% % Use the relevant command for your figure-insertion program
% % to insert the figure file.
% % For example, with the option graphics use
% \resizebox{0.75\textwidth}{!}{%
%   \includegraphics{leer.eps}
%\vspace{5cm}       % Give the correct figure height in cm
% \caption{Please write your figure caption here}
% \label{fig:1}       % Give a unique label
% \end{figure}
%\begin{figure*}
%\vspace*{5cm}       % Give the correct figure height in cm
% \caption{Please write your figure caption here}
% \label{fig:2}       % Give a unique label
% \end{figure*}
% \begin{table}
% \caption{Please write your table caption here}
% \label{tab:1}       % Give a unique label
% % For LaTeX tables use
% \begin{tabular}{lll}
% \hline\noalign{\smallskip}
% first & second & third  \\
% \noalign{\smallskip}\hline\noalign{\smallskip}
% number & number & number \\
% number & number & number \\
% \noalign{\smallskip}\hline
% \end{tabular}
% Or use
% \vspace*{5cm}  % with the correct table height
% \end{table}
%
% BibTeX users please use
%%\bibliographystyle{unsrt}
%%\bibliography{networks}

\end{document}